\documentclass[12pt]{article}
\pdfoutput=1
\usepackage{putex}
\usepackage{graphicx}
\usepackage{caption}
\usepackage{subcaption}
\usepackage{epstopdf}
\usepackage{mathtools}
\usepackage{enumerate}
\usepackage{cite}
\usepackage{tensor}
\usepackage{slashed}
\usepackage[aligntableaux=center]{ytableau}
\usepackage{multirow}

\numberwithin{equation}{section}

\newcommand{\abs}[1]{\left\lvert #1 \right\rvert}

\newcommand {\be} {\begin {equation}}
\newcommand {\ee} {\end {equation}}

\newcommand {\bes} {\begin {equation*}}
\newcommand {\ees} {\end {equation*}}

\newcommand{\es}[2] {\begin{equation} \label{#1} \begin{split} #2 \end{split} \end{equation}}

\newcommand{\Z}{\mathbb{Z}}

\def\<{\langle}
\def\>{\rangle}

\def\cO {{\cal O}}
\def\cG {{\cal G}}

\def\cN {{\cal N}}

\newcommand{\beq}{\begin{equation}}
\newcommand{\eeq}{\end{equation}}

\def\<{\langle}
\def\>{\rangle}

\begin{document}

\preprint{PUPT-2483}

\institution{PU}{Joseph Henry Laboratories, Princeton University, Princeton, NJ 08544}

\title{Accidental Symmetries and the\\ Conformal Bootstrap}

\authors{Shai M.~Chester, Simone Giombi, Luca V.~Iliesiu, \\[10pt] Igor R.~Klebanov, Silviu S.~Pufu, and Ran Yacoby}

\abstract{We study an $\cN = 2$ supersymmetric generalization of the three-dimensional critical $O(N)$ vector model that is described by $N+1$ chiral superfields with superpotential $W = g_1 X \sum_i Z_i^2 + g_2 X^3$.  By combining the tools of the conformal bootstrap with results obtained through supersymmetric localization, we argue that this model exhibits a symmetry enhancement at the infrared superconformal fixed point due to $g_2$ flowing to zero.  This example is special in that the existence of an infrared fixed point with $g_1,g_2\neq 0$, which does not exhibit symmetry enhancement, does not generally lead to any obvious unitarity violations or other inconsistencies. We do show, however, that the $F$-theorem excludes the models with $g_1,g_2\neq 0$ for $N>5$. The conformal bootstrap provides a stronger constraint and excludes such models for $N>2$. We provide evidence that the $g_2=0$ models, which have the enhanced $O(N)\times U(1)$ symmetry, come close to saturating the bootstrap bounds.  We extend our analysis to fractional dimensions where we can motivate the nonexistence of the $g_1,g_2\neq 0$ models by studying them perturbatively in the $4-\epsilon$ expansion.}
\date{}

\maketitle

\tableofcontents

\section{Introduction}

The technique of $F$-maximization, originally proposed in \cite{Jafferis:2010un} and further developed in \cite{Jafferis:2011zi, Closset:2012vg}, allows one to identify the superconformal $U(1)_R$ symmetry at the infrared fixed points of a wide class of three-dimensional RG trajectories that preserve ${\cal N} = 2$ supersymmetry. Identifying this $U(1)_R$ symmetry is desirable, since the scaling dimensions of BPS operators as well as unitarity bounds satisfied by operators that belong to long multiplets are all determined by their respective R-charges \cite{Minwalla:1997ka}.  Determining the R-symmetry of a superconformal field theory (SCFT) using $F$-maximization relies on embedding that SCFT as the infrared limit of an RG flow that starts from a free UV theory. The matter content of the UV theory and the symmetries preserved by the flow are used as inputs for the $F$-maximization procedure. The R-symmetry of the infrared fixed point is then determined in terms of these inputs; the explicit procedure is described in detail in \cite{Jafferis:2010un}.  This procedure can be carried out explicitly using the supersymmetric localization results of \cite{Jafferis:2010un, Hama:2010av}.

In applying the above method one implicitly assumes that the symmetries of the infrared SCFT are isomorphic to those preserved by the RG flow. An important caveat of $F$-maximization is that it may lead to incorrect results if this assumption fails. This may happen due to extra symmetries that emerge at the IR fixed point, but are not present throughout the flow; such extra symmetries are usually referred to as accidental.  Indeed, the superconformal R-symmetry at the IR fixed point could happen to be a linear combination of the various $U(1)$ symmetries preserved by the flow {\em and} of accidental $U(1)$ symmetries. In such cases the $F$-maximization procedure has to be modified in order to take those extra symmetries into account \cite{Niarchos:2011sn, Morita:2011cs, Agarwal:2012wd}.   One indication that accidental symmetries may be present is when $F$-maximization yields unitarity violating scaling dimensions for some chiral operators.\footnote{We refer the reader to \cite{Kutasov:2003iy} for a nice discussion and examples in the context of 4d $\cN=1$ gauge theories.} It is generally speculated that these chiral operators actually become free, resulting in the emergence of additional symmetries that were not taken into account. In some other cases, accidental symmetries can be detected by passing to a weakly coupled dual of the original theory, in which the symmetry enhancement is manifest \cite{Safdi:2012re}.  Finally, it is possible that the RG flows where the $F$-theorem \cite{Myers:2010xs, Jafferis:2011zi, Klebanov:2011gs, Casini:2012ei} seems to be violated could also indicate the presence of additional symmetries, but there is no known systematic way for detecting them.

Our work is focused on a simple family of models whose naive IR fixed points, obtained under the assumption of no symmetry enhancement, do not always exhibit any obvious problems such as those mentioned above. Nevertheless, we will show that these naive IR fixed points are ruled out as consistent unitary SCFTs, since their CFT data violates the constraints imposed by crossing symmetry. These constraints can be efficiently implemented with the help of the numerical conformal bootstrap technique introduced in \cite{Rattazzi:2008pe}. We view this result as strong evidence that accidental symmetries must be present in the IR\@. To our knowledge, our work represents the first example where the conformal bootstrap is used to argue for such a result.\footnote{The idea that the conformal bootstrap could be used to provide evidence for the emergence of accidental symmetries was already proposed in \cite{Poland:2010wg}.}

The theory we examine is a certain ${\cal N} = 2$ supersymmetric generalization of the critical $O(N)$ vector model.  This model has a chiral super-field $X$ and $N$ chiral super-fields $Z_i$ interacting through a cubic superpotential
 \es{SuperPot}{
  W = \frac {g_1}2  X \sum_{i=1}^N Z_i^2 + \frac {g_2}6 X^3 \,,
 }
where $g_1$ and $g_2$ are coupling constants.  The superpotential interaction triggers a supersymmetric RG flow that preserves $O(N) \times \Z_3$ flavor symmetry.  Under the $O(N)$ symmetry, the $Z_i$ transform as a fundamental vector and $X$ is a singlet.  The generator of the $\Z_3$ symmetry acts by multiplying both $X$ and $Z_i$ by $e^{2 \pi i /3}$.  A consequence of the $O(N) \times \Z_3$ symmetry is that no other superpotential interactions are dynamically generated throughout the flow.

The naive guess in this case is that generically, without further tuning of parameters, the RG trajectory would end at an IR fixed point preserving only $O(N) \times \Z_3$ flavor symmetry where both $g_1$ and $g_2$ are non-zero.\footnote{It should be clear that the couplings that flow are the ones obtained after rescaling $X$ and $Z_i$ such that their corresponding K\"{a}hler terms are canonical. For simplicity, we will not make a distinction between the superpotential and the canonical couplings in our notation.} At this fixed point, the superconformal $U(1)_R$ charges (and hence scaling dimensions) of $X$ and of $Z_i$ would all equal $2/3$, as this is the only R-symmetry preserved by the entire flow that commutes with $O(N) \times \Z_3$---in this case, $F$-maximization is rather trivial, since there is no freedom in what the superconformal R-symmetry could be.  Due to recent progress on studying supersymmetric field theories on curved manifolds, one could compute certain observables of this naive fixed point exactly.  For instance, one can use supersymmetric localization to compute its $S^3$ free energy $F = - \log \abs{Z_{S^3}}$,
or further using the results of \cite{Closset:2012vg}, one can also calculate the $O(N)$ ``central charge'' $c_J^{O(N)}$, defined in terms of the two-point function of the canonically normalized $O(N)$ current, or the coefficient $c_T$ defined in terms of the two-point function of the canonically normalized stress tensor.  The main argument of this paper is quite simple:  using the conformal bootstrap, we show that a unitary SCFT with $O(N)$ symmetry, a chiral $O(N)$ fundamental of dimension $2/3$, and with the value of $c_J^{O(N)}$ determined by localization, is inconsistent when $N>2$.\footnote{Using the value of $c_T$ of this fixed point by itself does not seem to lead to an inconsistency.} The naive fixed point of \eqref{SuperPot} therefore does not exist when $N>2$, and we expect symmetry enhancement in the infrared.  A weaker result can be obtained using the $F$-theorem \cite{Myers:2010xs, Jafferis:2011zi, Klebanov:2011gs, Casini:2012ei}, which can be used to argue that this naive fixed point does not exist when $N>5$.

The naive fixed point of \eqref{SuperPot} with non-vanishing $g_1$ and $g_2$ also does not exist for $N=2$.  In this case, if $g_2 = 0$, the theory is equivalent to the $XYZ$ model.  The $X^3$ coupling in \eqref{SuperPot} is then marginal, and using conformal perturbation theory around the $XYZ$ model, it can be seen to flow to zero in the infrared.\footnote{In fact, in 3d $\cN=2$ SCFTs, a marginal operator can be either exactly marginal or marginally irrelevant \cite{Green:2010da}. It is known that the $XYZ$ model has one exactly marginal deformation, and that it preserves the permutation symmetry between $X$, $Y$, and $Z$ \cite{Strassler:1998iz}. The marginal deformation $X^3$ does not preserve this symmetry and must therefore be marginally irrelevant.}   The assumptions used in our particular bootstrap analysis do not distinguish between the $XYZ$ model and its $X^3$ deformation as in \eqref{SuperPot}, which is why we do not exclude the $N=2$ case. When $N=1$, we believe that generically \eqref{SuperPot} flows to two decoupled copies of the ${\cal N} =2$ super-Ising theory (the latter was studied in a bootstrap context in \cite{Bobev:2015jxa}).  We will give some evidence for this proposal in Section~\ref{PERTURBATIVE}.   Our main focus in this paper is, however, the case $N>2$.

When $N>2$, what we believe happens with the model \eqref{SuperPot} is that the IR physics is governed by one of two other fixed points with enhanced symmetry.  Apart from the UV fixed point with $g_1 = g_2 = 0$, the theory \eqref{SuperPot} is also believed to have a fixed point where $g_2 = 0$ and $g_1 \neq 0$, as well as a fixed point where $g_1 = 0$ and $g_2 \neq 0$. The former has $O(N) \times U(1)$ flavor symmetry, as we review in Section~\ref{RG}.  The latter fixed point has $U(N) \times \Z_3$ flavor symmetry (the $\Z_3$ is as before and under $U(N)$ the $Z_i$ transform as a fundamental and $X$ is a singlet), being the product of a free theory of $N$ chiral multiplets and the ${\cal N} = 2$ super-Ising model studied in \cite{Bobev:2015jxa}.   Both fixed points enhance the $O(N) \times \Z_3$ preserved by the flow to a strictly larger symmetry group.

Among the applications of the models \eqref{SuperPot} with $N>2$ are the world volume theories of M2-branes placed at Calabi-Yau singularities.
For example, the theory with $N=5$ has been proposed in \cite{Martelli:2009ga}\footnote{At first sight, the theory proposed in \cite{Martelli:2009ga} seems different because it is a $U(1)\times U(1)$ Chern-Simons gauge theory.
However, for $k=1$ the gauge symmetries become unimportant, and the model reduces to the $SO(5)$ symmetric Wess-Zumino model \eqref{SuperPot} after some field redefinitions.} as a description of a single M2-brane placed at the tip of the conical Stenzel space
$\sum_{i=1}^5 z_i^2=0$.
This claim is plausible because the classical moduli space of the model \eqref{SuperPot} is $X=0$, $\sum_{i=1}^5 Z_i^2=0$, which is exactly the conical Stenzel space.
We will show, however, that the $SO(5)$ symmetric theory with non-vanishing $g_1$ and $g_2$ does not exist at the quantum level. We postpone a further discussion of models that arise on M2-branes to future work.

The rest of this paper is organized as follows.  In Section~\ref{INTUITION}, we provide some Renormalization Group arguments for why for $N>2$  one does not expect a fixed point of \eqref{SuperPot} with both $g_1$ and $g_2$ non-vanishing. In particular, the $F$-theorem rules out such models for $N>5$. In Section~\ref{BOOTSTRAP}, we provide rigorous bounds on $c_J^{O(N)}$ when the R-charge of $Z_i$ equals $2/3$ and notice that the naive fixed point of \eqref{SuperPot} would be inconsistent for $N>2$.  In this section, we actually present our bounds more generally in $d$ space-time dimensions, with $2<d<4$.  We leave the computation of $c_J^{O(N)}$ using supersymmetric localization, as well as a detailed description of the bootstrap equations to the Appendices.

\section{Renormalization Group Arguments}
\label{INTUITION}

In this section we present some arguments that, for $N > 2$, the theory \eqref{SuperPot} is not expected to have a fixed point with only $O(N) \times \Z_3$ flavor symmetry. The first argument relies on a general RG flow analysis and the second on computations done in the $4-\epsilon$ expansion. The third argument uses the $F$-theorem with which we are able to exclude our putative SCFT, but only for $N>5$. A more rigorous proof that this putative CFT does not exist for all $N>2$ will be given in Section \ref{BOOTSTRAP} using the conformal bootstrap.

\subsection{RG Flow Analysis}
 \label{RG}

Let us start by analyzing the predictions of supersymmetry for various flows in the space of couplings $g_1$ and $g_2$ of our model \eqref{SuperPot}. Consider first deforming the free theory $g_1=g_2=0$ only by the relevant deformation $X^3$, i.e., turning on $g_2\neq 0$, while keeping $g_1=0$ in the superpotential \eqref{SuperPot}. The theory flows to the super-Ising SCFT plus $N$ free fields $Z_i$, in which the R-charges of $X$ and $Z_i$, are given by $r_Z = \frac{1}{2}$ and $r_X=\frac{2}{3}$, respectively. In this case, these charges are fixed trivially by demanding the marginality of the superpotential and the fact that the $Z_i$ are free. The deformation $X\sum Z_i^2$ has R-charge $r_{X Z^2} = \frac{2}{3} + 2\times \frac{1}{2} < 2$, and so it is a relevant deformation of the super-Ising plus $N$ free fields SCFT\@. Turning on a non-zero $g_1$ we therefore expect to flow to a non-trivial SCFT in the IR in which $r_X=r_Z=\frac{2}{3}$. It is easy to see that this series of RG flows is consistent with the $F$-theorem.

The second series of flows is slightly more involved, and starts by considering the case $g_2 = 0$ in \eqref{SuperPot}, where the superpotential becomes \cite{Strassler:2003qg,Nishioka:2013gza}
 \es{SuperPotEnhanced}{
  W =  \frac {g_1}2  X \sum_{i=1}^N Z_i^2 \,.
 }
This superpotential preserves more flavor symmetry than \eqref{SuperPot}:  it preserves an $O(N)$ symmetry under which the $Z_i$ transform as a fundamental vector and $X$ is a singlet, as well as an additional $U(1)$ under which $Z_i$ and $X$ have charges $+1$ and $-2$, respectively.  Since \eqref{SuperPotEnhanced} is the most general superpotential interaction preserving this $O(N) \times U(1)$ flavor symmetry, no other superpotential interactions are generated throughout the RG flow.

The infrared limit of \eqref{SuperPotEnhanced} is believed to be an ${\cal N} = 2$ superconformal field theory analog of the critical $O(N)$ model.  The scaling dimensions of the chiral operators $X$ and $Z_i$ are related to their superconformal R-charges, which can be determined using $F$-maximization and supersymmetric localization as in \cite{Nishioka:2013gza}. In this case, the most general R-symmetry preserved by \eqref{SuperPotEnhanced} is such that the R-charges of $X$ and $Z_i$ obey
 \es{rXZRelation}{
   r_X = 2 - 2 r_Z \,,
 }
with arbitrary $r_Z$.  (The superpotential has R-charge $2$.)  Maximizing the $S^3$ free energy over $r_Z$, Ref.~\cite{Nishioka:2013gza} obtained the values listed in Table~\ref{RCharges} for various values of $N$.  The superconformal R-charges, and thus scaling dimensions of the operators of this SCFT, are thus quite non-trivial.  For future reference, in this table we also list the values of $c_J^{O(N)}$ for this SCFT, computed using the method explained in Appendix~\ref{LOCALIZATION}.
\begin{table}[htdp]

\begin{center}
  \begin{tabular}{c|c|c|c|c|c|c|c|c|c|c}
   $N$ & $1$ & $2$ & $3$ & $4$ & $5$ & $6$ & $7$ & $8$ & $9$ & $10$  \\
 \hline\hline
 $r_Z$ & $.708$ & $.667$ & $.632$ & $.605$ & $.586$ & $.572$ & $.562$ & $.554$ & $.548$ & $.543$ \\
 \hline
 $r_X$ & $.584$ & $.667$ & $.737$ & $.790$ & $.828$ & $.856$ & $.876$ & $.892$ & $.904$ & $.914$ \\
 \hline
 $c_J^{O(N)}$ & -- & $.521$ & $.600$ & $.664$ & $.715$ & $.754$ & $785$ & $809$ & $.828$ & $.844$
  \end{tabular}
\end{center}
\caption{The superconformal R-charges $r_X$ and $r_Z$ of $X$ and $Z_i$, respectively, as well as the coefficient $c_J^{O(N)}$ at the infrared fixed point of \eqref{SuperPotEnhanced}.  The coefficient $c_J^{O(N)}$ is normalized so that it equals $1$ in a theory of $N$ free chiral multiplets. }
\label{RCharges}
\end{table}%

Coming back to the theory \eqref{SuperPot} (with $g_1,g_2\neq 0$), one may interpret it as a superpotential deformation of the IR fixed point of \eqref{SuperPotEnhanced} by the operator $X^3$.  One can see, however, that if $N>2$, at the fixed point of \eqref{SuperPotEnhanced}, the operator $X^3$ has R-charge $r_{X^3} > 2$, and so the superpotential deformation by it is irrelevant.  In other words, if one starts at the interacting superconformal fixed point that has $g_1 \neq 0$ and $g_2 = 0$ and turns on a small non-zero value for $g_2$, then $g_2$ flows back to zero.  An ${\cal N} = 2$ fixed point with non-zero $g_1$ and $g_2$, if it exists at all, can therefore not be reached from the fixed point with $g_2 = 0$.  Moreover, the arguments of \cite{Green:2010da} guarantee that if such a fixed point exists as a unitary SCFT, it is attractive in the space of couplings $(g_1, g_2)$.

To summarize, we naively expect the space of theories described by \eqref{SuperPot} to contain four types of fixed points. The free theory in which both $g_1$ and $g_2$ are relevant; the super-Ising model plus $N$ free fields that preserves $U(N)\times \mathbb{Z}_3$ symmetry and in which $g_1$ is relevant, but $g_2$ is irrelevant; the model \eqref{SuperPotEnhanced} that preserves $O(N)\times U(1)$ symmetry and in which both $g_1$ and $g_2$ are irrelevant (for $N>1$); and a fixed point preserving only $O(N)\times\mathbb{Z}_3$ where both $g_1$ and $g_2$ are irrelevant.  The resulting RG flow diagram is quite peculiar.  The fixed point with only $O(N) \times \mathbb{Z}_3$ symmetry looks like a deformation of the one with $O(N) \times U(1)$ symmetry, but we have argued that it is impossible to have an RG flow line connecting the two, in either direction.  It is thus reasonable to expect that one of these two fixed points does not correspond to a unitary SCFT\@.  In the next section we will show that it is the $O(N) \times \mathbb{Z}_3$ one that does not exist.

\subsection{$4-\epsilon$ Expansion}
\label{PERTURBATIVE}

\begin{figure}[ht!]
\begin{center}
 \includegraphics[width = 0.49\textwidth]{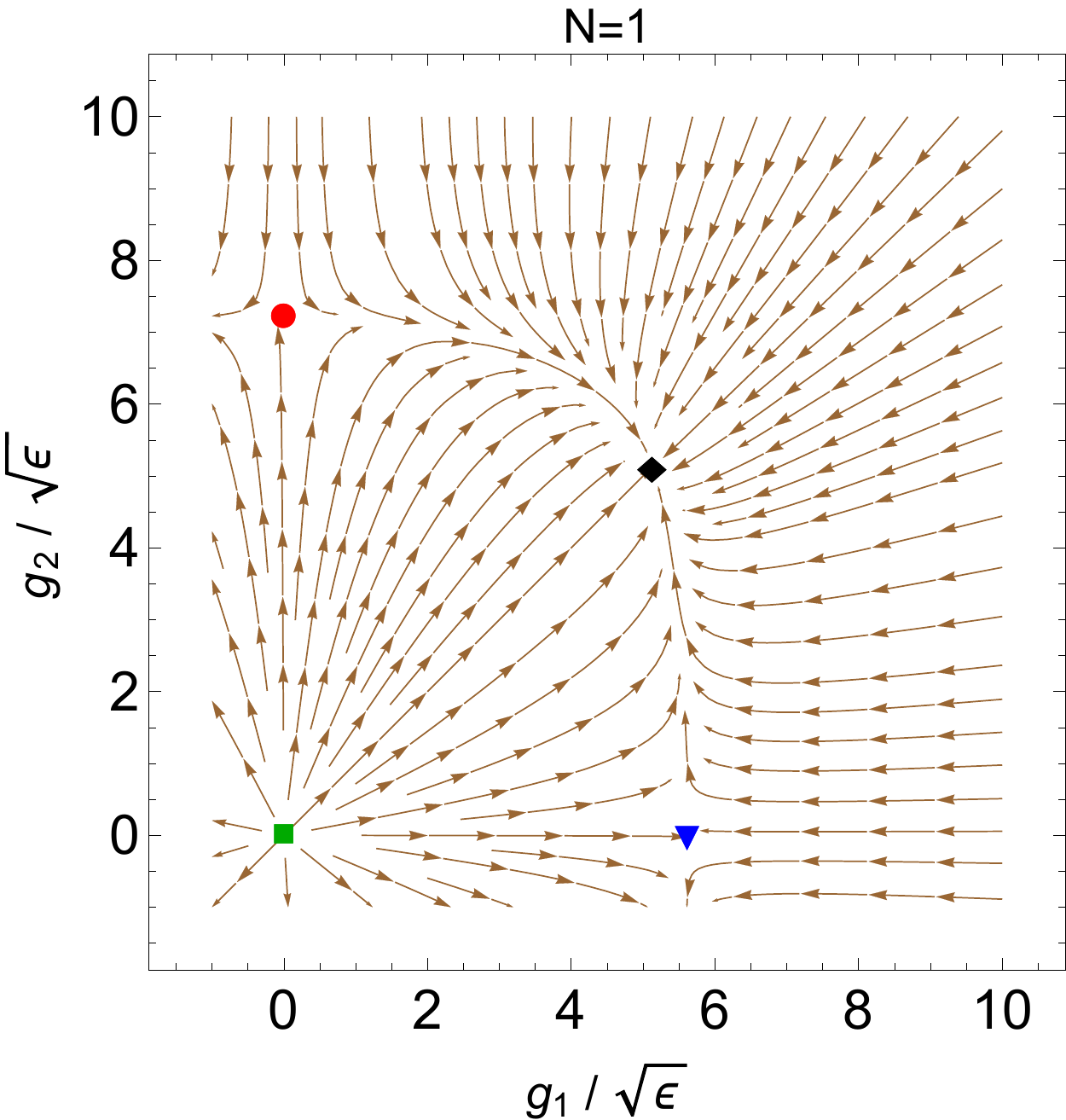}
 \includegraphics[width = 0.49\textwidth]{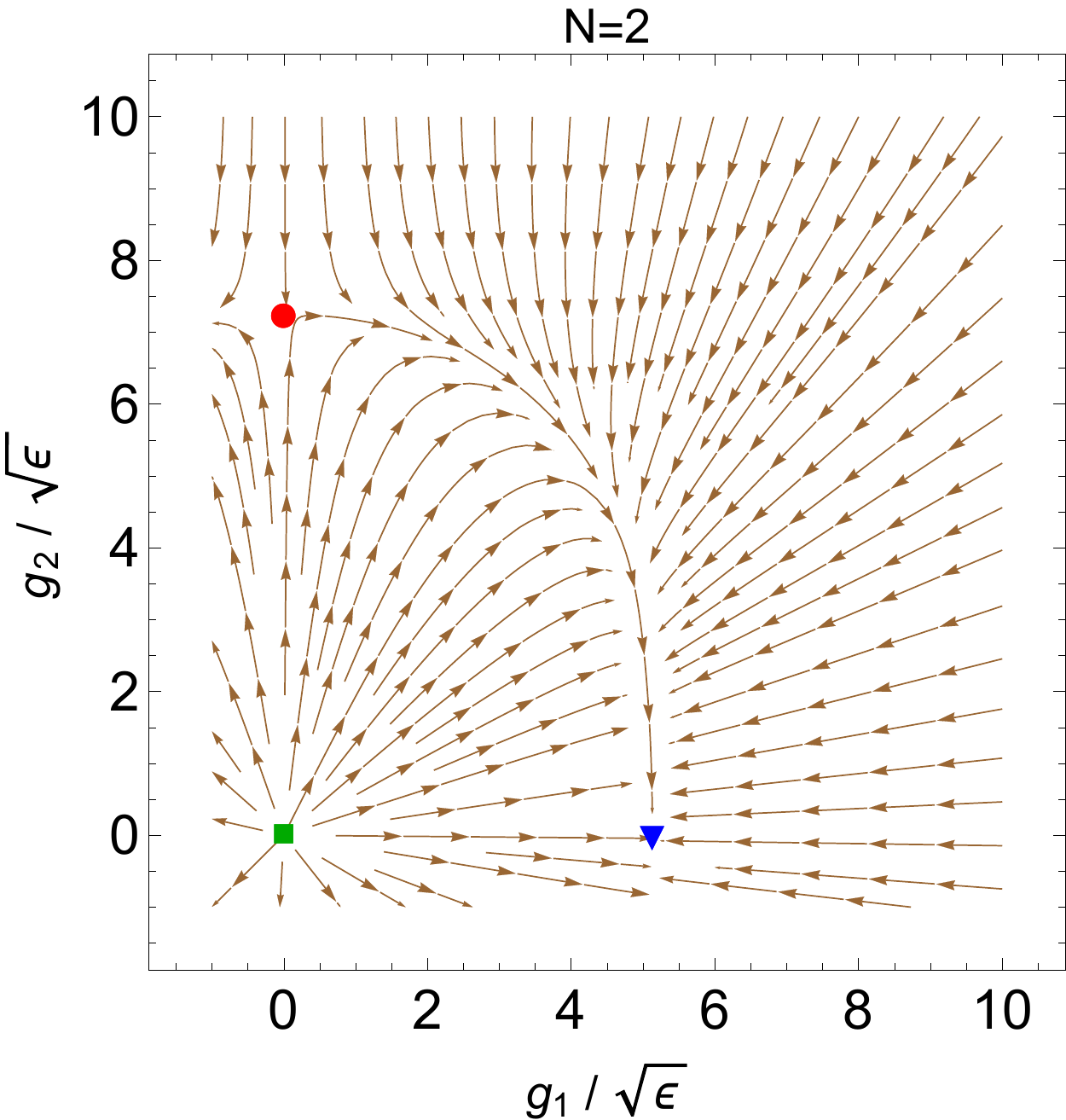} \\
 \includegraphics[width = 0.49\textwidth]{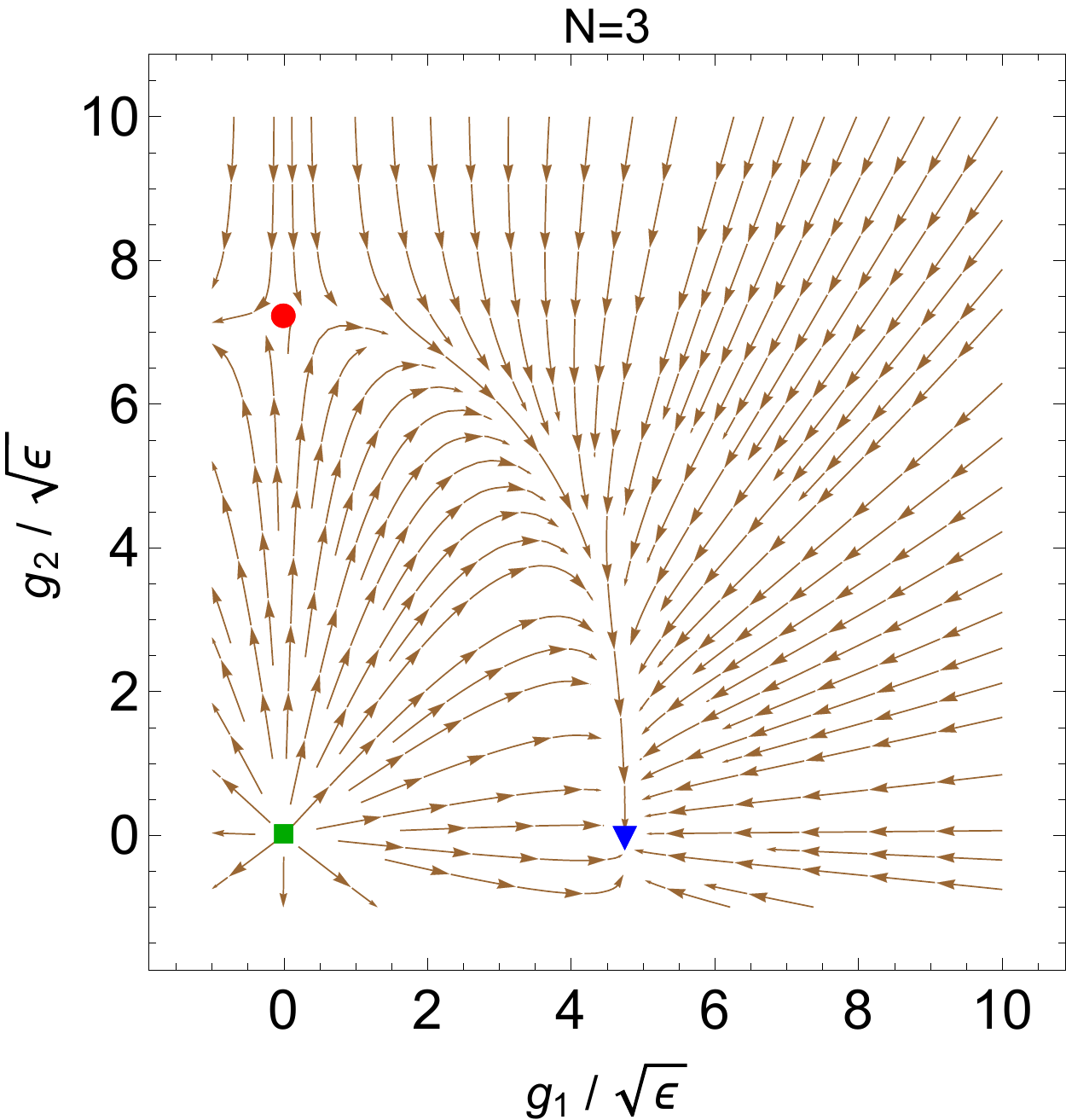}
 \caption{RG flow lines obtained from the one-loop beta functions---see \eqref{beta}, where these beta functions are given to two-loop order.  The green square, red circle, blue triangle, and black diamond correspond to the fixed points in \eqref{Solutions}. Note that the last fixed point only exists for $N=1$.  \label{fig:StreamPlots}}
 \end{center}
 \end{figure}

Our second argument why for $N>2$ we do not expect a fixed point with only $O(N) \times \Z_3$ flavor symmetry comes from studying the models \eqref{SuperPot} in the $4-\epsilon$ expansion.  From the results of \cite{Ferreira:1996az}, we can read off the beta functions for the physical couplings in $4-\epsilon$ dimensions.\footnote{The physical couplings are $g_1^\text{phys} = g_1 / \sqrt{Z_X Z_Z^2}$ and $g_2^\text{phys} = g_2 / \sqrt{Z_X^3}$, where $Z_X$ and $Z_Z$ are the wave-function renormalization factors for $X$ and $Z$, respectively, in a holomorphic scheme where the superpotential is not renormalized. Above, we use the notation $g_i$ instead of $g_i^\text{phys}$ to avoid clutter.}
They are known up to four loops, but here we state them up to the two-loop order for brevity:
 \es{beta}{
  \beta_1 &= g_1 \left[- \frac{\epsilon}{2} + \frac{(N+4) \abs{g_1}^2 + \abs{g_2}^2 }{32 \pi^2 } -
  \frac{ 4 (N+1) \abs{g_1}^4 + (N+2) \abs{g_1}^2 \abs{g_2}^2+ \abs{g_2}^4 }{512 \pi^4 }+ \ldots
  \right] \,, \\
  \beta_2 &= g_2 \left[-\frac{\epsilon}{2} + \frac{3 \left( N \abs{g_1}^2 +  \abs{g_2}^2 \right) }{32 \pi^2}
  -\frac{3( 2N \abs{g_1}^4 + N \abs {g_1}^2 \abs{g_2}^2+ \abs{g_2}^4  ) }{512 \pi^4 }
    + \ldots \right] \,.
 }
The equations $\beta_1 = \beta_2 = 0$ have the following perturbative solutions:
 \es{Solutions}{
  I&: \quad g_1 = g_2 = 0 \,, \\
  II&: \quad g_1 = 0 \,, \qquad \abs{g_2}^2  = \frac{16 \pi^2\epsilon}{3} \left(1+\frac{\epsilon}{3}
+\left(\frac{1}{12}-\frac{\zeta(3)}{3}\right)\epsilon^2+O(\epsilon^{3}) \right)\,,\\
  III&: \quad g_2 = 0 \,,\\
& \text{\hspace{0.75cm}} \abs{g_1}^2 =
\frac{16 \pi^2\epsilon}{N+4}\left(1+\frac{4(N+1)}{(N+4)^3}\epsilon
+\frac{16+16N+17N^2-N^3-6(N+4)^2\zeta(3)}{(N+4)^4}\epsilon^2+O(\epsilon^{3})\right)\,,\\
  IV&: \quad \abs{g_1}^2 = \frac{8 \pi^2\epsilon}{3} \left (1  +\frac {\epsilon}{3}
+\frac{1+2(N-3)\zeta(3)}{12}\epsilon^2+ O(\epsilon^{3})\right )\,, \qquad\\
  & \text{\hspace{0.75cm}} \abs{g_2}^2 = \frac{8 \pi^2 \epsilon}{3} (2-N) \left (1 +\frac {\epsilon}{3}+
\frac{1-(N^2-N+4)\zeta(3)}{12}\epsilon^2+O(\epsilon^{3})\right ) \,.
}
These equations determine $g_1$ and $g_2$ up to arbitrary phases that can be absorbed through field redefinitions.  See Figure~\ref{fig:StreamPlots} for plots of the RG flow lines obtained from the one-loop beta functions.  In this figure, the fixed points listed above are marked with a green square, a red circle, a blue triangle, and a black diamond, respectively.

The first solution is the free UV fixed point of theory \eqref{SuperPot}.  The second has $U(N) \times \Z_3$ flavor symmetry corresponding to a product SCFT of the supersymmetric Ising model and $N$ free chiral multiplets.  The third fixed point has $O(N) \times U(1)$ flavor symmetry.  The last fixed point only exists for $N=1$ (when $N=2$ it coincides with the third fixed point).  It corresponds to two decoupled copies of the super-Ising theory, as can be seen by making the field redefinitions $X \to  \frac{1}{\sqrt{2}}(X+Z)$ and $Z \to \frac{1}{\sqrt{2}}(X-Z)$. These redefinitions lead to two decoupled copies of the super-Ising theory whenever $|g_1|=|g_2|$, and one can indeed verify that this equality is satisfied by the fourth solution in \eqref{Solutions}.

We see that there is no fixed point that has $O(N) \times \Z_3$ global symmetry perturbatively in the $4 - \epsilon$ expansion.  It is plausible, however, that such a fixed point would appear non-perturbatively, so the $4 - \epsilon$ expansion cannot be used to argue conclusively that such a fixed point would also be absent when $\epsilon = 1$.

\subsection{$F$-theorem Arguments}
\label{Farguments}

The $F$-value of the naive fixed point of \eqref{SuperPot}, where both $g_1$ and $g_2$
 are non-vanishing, can be computed using the supersymmetric localization results of \cite{Jafferis:2010un}.  It equals $-(N+1)\ell(1/3)$, where $\ell(z)$ is the function defined in \cite{Jafferis:2010un} representing the contribution to the $S^3$ free energy of a chiral multiplet of R-charge $1 - z$, since at the naive fixed point of \eqref{SuperPot} the fields $X$ and $Z_i$ all have R-charge $2/3$.  Numerically, we have
 \es{FFixed}{
  F_{O(N)\times \Z_3} \approx 0.291 (N+1) \,.
 }
One could imagine deforming this fixed point by giving the $X$ field a large superpotential mass, $m_X X^2$.  As can be seen from solving the classical equation of motion, integrating out $X$ then induces a quartic superpotential interaction $(Z_i Z_i)^2$ for the $Z_i$ fields.  If we then fine-tune the $O(N)$-invariant mass of the $Z_i$ fields (a smaller and smaller degree of fine-tuning is required as we take $m_X  \to \infty$), the IR limit can be argued to be simply a free theory of the $N$ fields $Z_i$.  Its $F$-coefficient is
 \es{FIR}{
  F_\text{$N$ free fields} = N \frac{\log 2}{2} \approx 0.347 N \,.
 }
The RG flow we described must be possible if the fixed point with $O(N) \times \Z_3$ flavor symmetry exists.  We see, however, that $F_{O(N)\times \Z_3} < F_\text{$N$ free fields}$ when $N>5$, contradicting the $F$-theorem \cite{Myers:2010xs, Jafferis:2011zi, Klebanov:2011gs, Casini:2012ei}.  This $F$-theorem argument thus rules out the existence of a fixed point of \eqref{SuperPot} with non-vanishing $g_1$ and $g_2$ for $N>5$.  This is a weaker bound than the one obtained using the conformal bootstrap in the next section.

The theory with $g_2=0$, (\ref{SuperPotEnhanced}), may also be deformed by the relevant operator $X^2$.
This flow was considered in \cite{Nishioka:2013gza}
and shown to also lead to a free theory of $N$ chiral superfields $Z_i$. While such a flow provides
a counter-example to the $c_T$ theorem, the $F$-theorem holds for all $N$ \cite{Nishioka:2013gza}.
This is consistent with our arguments: the enhanced symmetry theory with $g_2=0$ exists for all $N$.

\begin{figure}[t]
\begin{center}
 \includegraphics[width = 0.6\textwidth]{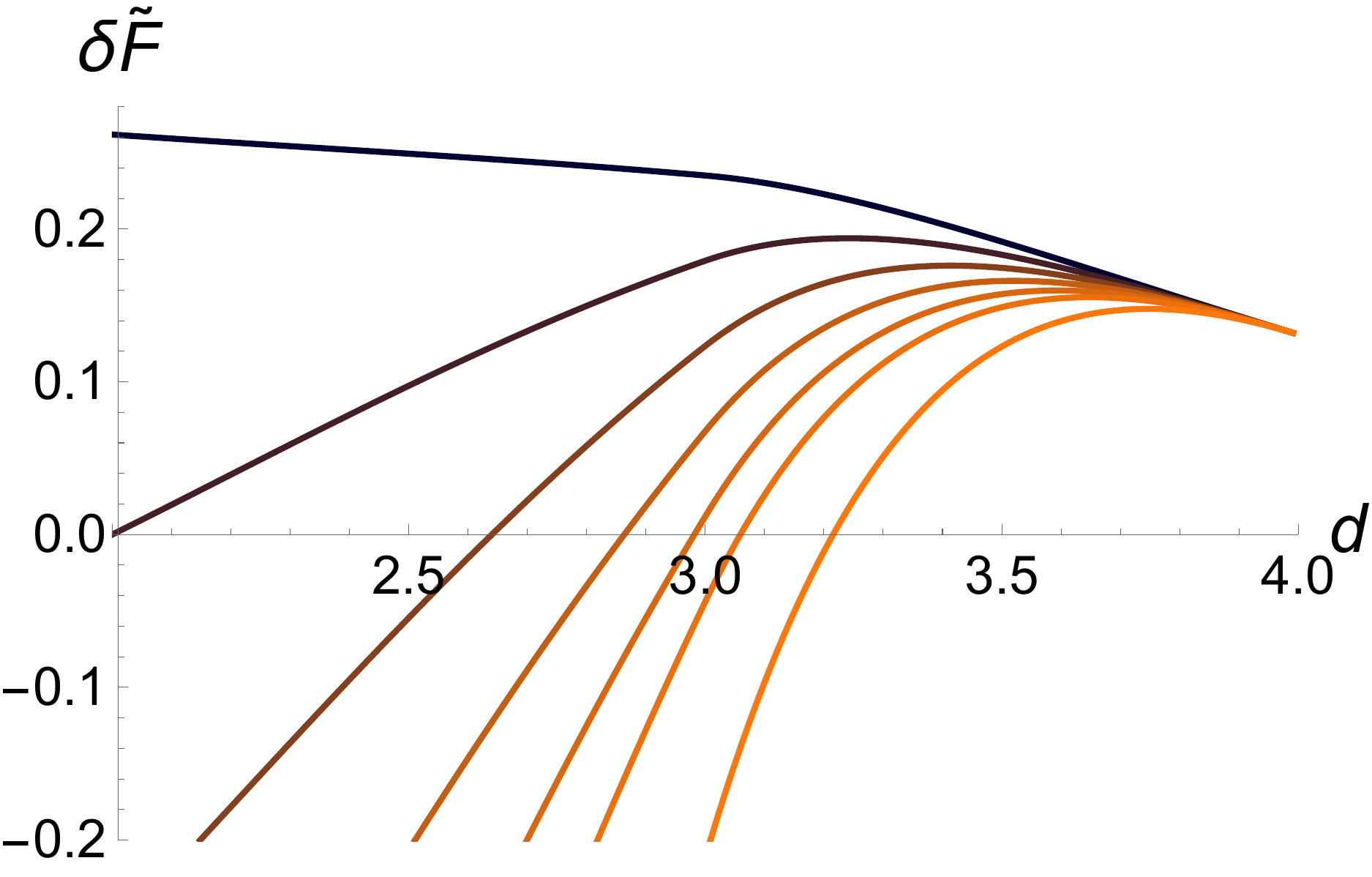}
 \caption{ $\delta{\tilde{F}}\equiv \tilde{F}_{UV} - \tilde{F}_{IR}$ as a function of $2< d < 4$ for $N=1,2,3,4,5,6,9$ (dark to light). Here, $\tilde{F}_{UV}$ is the generalized free energy defined in \cite{Giombi:2014xxa} corresponding to the theory \eqref{SuperPot} with $O(N)\times \Z_3$ symmetry, while $\tilde{F}_{IR}$ corresponds to the infrared fixed point obtained by deforming that theory by $X^2$. \label{fig:Fmax}}
 \end{center}
\end{figure}

It has been conjectured that the three-dimensional $F$-theorem is a special case of
the Generalized $F$-Theorem valid in continuous dimension
 \cite{Giombi:2014xxa,Fei:2015oha}. The conjecture states that $\tilde F_{\rm UV}>\tilde F_{IR}$,
 and for Wess-Zumino theories with 4 supercharges
 \begin{equation}
\tilde F = \sum_{\rm chirals} \tilde {\cal F}(\Delta_i)\ ,
\label{exact-F}
\end{equation}
where the function ${\cal F}(\Delta)$ is given in eq. (5.23) of \cite{Giombi:2014xxa}.
In Figure \ref{fig:Fmax} we exhibit the constraints on allowed values of $N$ from the Generalized $F$-Theorem.
  They arise when the theory (\ref{SuperPot}) in continuous dimension $d$ is deformed by the operator $X^2$. For $3\leq d<4$ it flows to the free theory of $N$ chiral superfields, while for
   $2<d<3$ it flows to the theory with superpotential $(Z_i Z_i)^2$ and $\Delta_Z = (d-1)/4$. We see that the requirement
 $\tilde F_{\rm UV}>\tilde F_{IR}$ translates into increasingly stringent constraints on $N$ as $d$ decreases. For dimensions slightly above $d=2$
 all theories with $N>2$ are ruled out by  the Generalized $F$-Theorem. Directly in $d=2$ none of the theories (\ref{SuperPot}) with $N>1$ are expected to exist on general grounds, since there
 should not be such interacting CFTs with continuous symmetries acting on scalar fields.
 The $N=1$ theory is known to exist in $d=2$---it is a member of the $D_k$ series of ${\cal N}=2$ superconformal minimal models \cite{Vafa:1988uu,Martinec:1988zu}.

\section{Excluding Theories via Conformal Bootstrap}
\label{BOOTSTRAP}

We now aim to provide a more rigorous argument why an ${\cal N} = 2$-preserving fixed point of \eqref{SuperPot} with only $O(N) \times \Z_3$ flavor symmetry is inconsistent as a unitary theory.  This argument is non-perturbative and combines the conformal bootstrap with supersymmetric localization results.  The conformal bootstrap technique can be used to numerically bound scaling dimensions and OPE coefficients of low-lying operators of all unitary CFTs with a given global symmetry \cite{Rattazzi:2008pe,Rattazzi:2010yc, ElShowk:2012ht, El-Showk:2014dwa, Vichi:2011ux, Kos:2013tga,Chester:2014gqa,Nakayama:2014lva,Nakayama:2014sba} or a certain amount of supersymmetry \cite{Poland:2010wg, Poland:2011ey,Beem:2013qxa, Beem:2013hha,Alday:2013opa, Alday:2013bha,Chester:2014fya,Chester:2014mea}. At the same time, scaling dimensions and OPE coefficients of operators protected by supersymmetry can be determined analytically for a given theory using supersymmetric localization \cite{Nishioka:2013gza}. If the bootstrap bounds exclude the values determined via localization, then such theories cannot be consistent unitary SCFTs.

The ``central charge'' of the $O(N)$ conserved current $c_J^{O(N)}$ for $\mathcal{N}=2$ SCFTs is an example of a quantity that can both be bounded using the bootstrap and computed exactly from supersymmetric localization.  In $d$ space-time dimensions, we define the central charge $c_J^{O(N)}$ of the canonically normalized $O(N)$  conserved current $j^\mu_{ij}$ by
\es{ONcurrent}{
  \langle j^\mu_{ij}(x) j^\nu_{kl}(0) \rangle = c_J^{O(N)}\frac{\Gamma^2(d/2)}{4(d-1)(d-2)\pi^d} \left(\delta_{ik}\delta_{jl}-\delta_{il}\delta_{jk}\right)
      \left(\eta^{\mu\nu}-2\frac{x^\mu x^\nu}{x^2}\right)\frac{1}{x^{2d-2}} \,.
}
With this definition, $c_J^{O(N)} = 1$ for a free chiral multiplet transforming in the fundamental representation of $O(N)$.
(While we are primarily interested in the theory \eqref{SuperPot} defined in three space-time dimensions, we will be more general and perform a study in $d$ space-time dimensions.)

In Appendix~\ref{LOCALIZATION} we explain how to compute $c_J^{O(N)}$ using supersymmetric localization.  For the proposed fixed point of \eqref{SuperPot} with only $O(N) \times \Z_3$ flavor symmetry, we have that the R-charges of $X$ and $Z$ are fixed to be $r_X = r_Z = 2/3$ in any $d$, but the relation between the scaling dimension and R-charge is dimension dependent \cite{Giombi:2014xxa,Bobev:2015jxa}:
\es{zGend}{
\Delta_{Z_i}=\Delta_X=\frac{d-1}{3} \,.
}
The result for $c_J^{O(N)}$ is also a function of $d$ but independent of $N$.  For instance, in $d=3$, we have
 \es{cJ3d}{
   c_J^{O(N)} =  \frac 89 - \frac {2}{\pi \sqrt{3}} \approx 0.521 \,,
 }
in agreement with the value listed in Table~\ref{RCharges} for $N=2$, in which case the theory \eqref{SuperPotEnhanced} also has $r_X = r_Z = 2/3$.  For a list of values for $c_J^{O(N)}$ in various space-time dimensions, see Table~\ref{cJONTable}.

\begin{table}[htdp]
\begin{center} \begin{tabular}{c|c|c|c|c|c|c|c|c|c|c|c|c|c|c|c|c|c|c}
 $d$ & 2.2&2.4&2.6&2.8&3&3.2&3.4&3.6&3.8&4 \\
 \hline
 $c_J^{O(N)}$ &.050&.154&.275&.399&.521&.637&.744&.841&.927&1
  \end{tabular}
  \end{center}
  \caption{The values of $c_J^{O(N)}$ for a potential fixed point of \eqref{SuperPot} with only $O(N) \times \Z_3$ flavor symmetry.}
  \label{cJONTable}
 \end{table}

  \begin{figure}[t!]
\begin{center}
 \includegraphics[width = 0.7\textwidth]{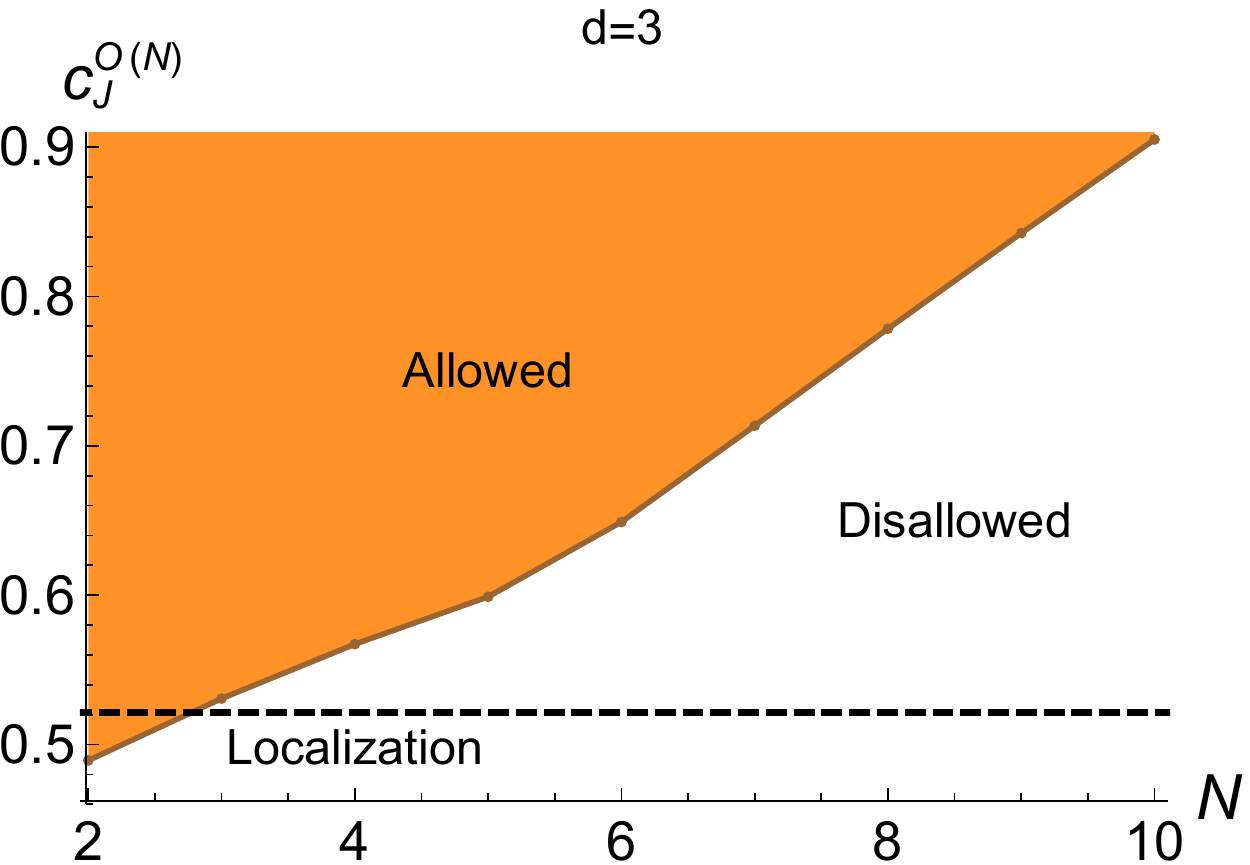}
 \caption{Lower bounds on central charge $c_{J}^{O(N)}$ for $\mathcal{N}=2$ SCFTs with $O(N)$ symmetry in $d=3$ for $N=2,...,10$, computed using the conformal bootstrap. The black dotted line denotes the $N$ independent analytical value of $c_{J}^{O(N)}$ computed  from localization for SCFTs with super potential \eqref{SuperPot}. For $N\geq 3$, the  black dotted line falls outside of the region allowed by the bootstrap, making model \eqref{SuperPot} a disallowed SCFT.   \label{fig:BoundN}}
 \end{center}
 \end{figure}

Since the $O(N)$ current $j^\mu_{ij}$ is a superconformal descendant in the same super-multiplet as the spin-0 superconformal primary $\cO_{Sa,d-2,0}$, we can relate $c_J^{O(N)}$ to the OPE coefficient $\lambda_{Sa,d-2,0}$. (Our convention for denoting operators is ${\cal O}_{X x, \Delta, \ell}$ where $X = S, T$ corresponds to $U(1)_R$ charge $0$ and $\pm 2 r_Z$, respectively; $x = s, t, a$ corresponds to the singlet, rank-two symmetric traceless, and antisymmetric $O(N)$ representation;  $\Delta$ is the scaling dimension; and $\ell$ is the spin.  We use a similar convention for OPE coefficients.)  In our normalization,
  \es{ONcJLambda}{
 c_J^{O(N)}=\frac{2^{2d-5}}{\lambda^2_{Sa,d-2,0}} \,.
 }
As explained further in Appendix~\ref{CROSSING}, we can now use the conformal bootstrap to place upper bounds on $\lambda_{Sa,d-2,0}^2$, and therefore lower bounds on $c_J^{O(N)}$, as a function of $N$, space-time dimension $d$, and the scaling dimension of the superconformal primary $\Delta_{Z_i}$.

 \begin{figure}[t!]
 \begin{center}
  \includegraphics[width = 0.7\textwidth]{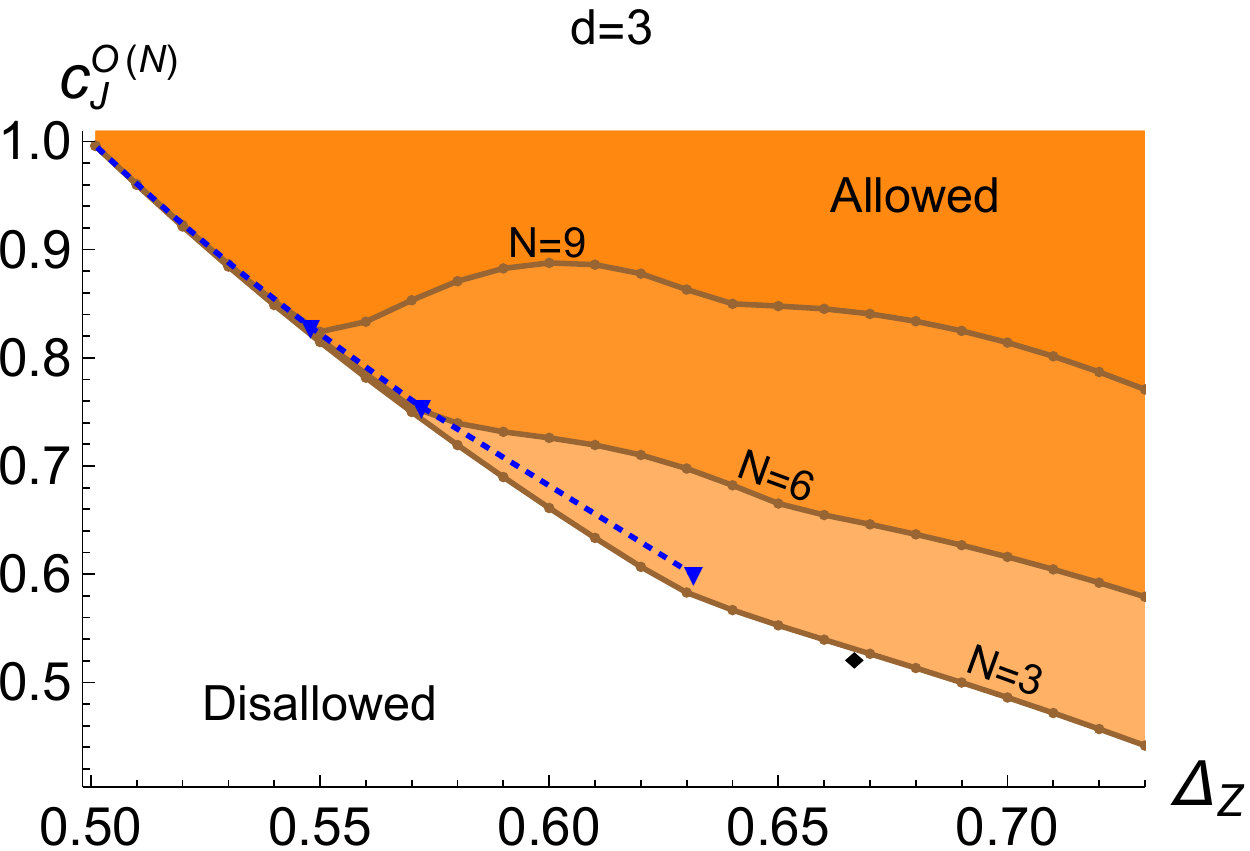}
 \caption{Conformal bootstrap lower bounds on $c_{J}^{O(N)}$ for $\mathcal{N}=2$ SCFTs with $O(N)$ symmetry in 3d, for $N=3, 6, 9$. The black diamond denotes the $N$-independent analytical value of $(\Delta_{Z_i},c_{J}^{O(N)})=(2/3,0.521)$, computed from localization for SCFTs with super potential \eqref{SuperPot}. The diamond falls outside of the orange shaded region allowed by the bootstrap.  The blue triangles and the dotted blue line correspond to the interacting SCFT with $O(N) \times U(1)$ flavor symmetry---see Table~\ref{RCharges}.  The blue triangles correspond to $N=3, 6, 9$, from right to left.}
 \label{fig:cJvsDeltaZ}
 \end{center}
 \end{figure}

In Figure~\ref{fig:BoundN} we show lower bounds on $c_J^{O(N)}$ at the value $\Delta_{Z_i}=\frac{2}{3}$ determined by R-Symmetry for $N=2,...,10$ and $d=3$, along with the localization value $c_J^{O(N)}\approx 0.521$. This value is disallowed by the bootstrap bounds for $N\geq 3$ and thus shows that the IR limit of the model \eqref{SuperPot} cannot be described by an $\cN=2$  SCFT with only $O(N) \times \Z_3$ flavor symmetry.  We view this as a non-perturbative proof that for $N \geq 3$ the above flow must exhibit flavor symmetry enhancement in the IR.

  \begin{figure}[t!]
   \begin{center}
 \includegraphics[width = 0.49\textwidth]{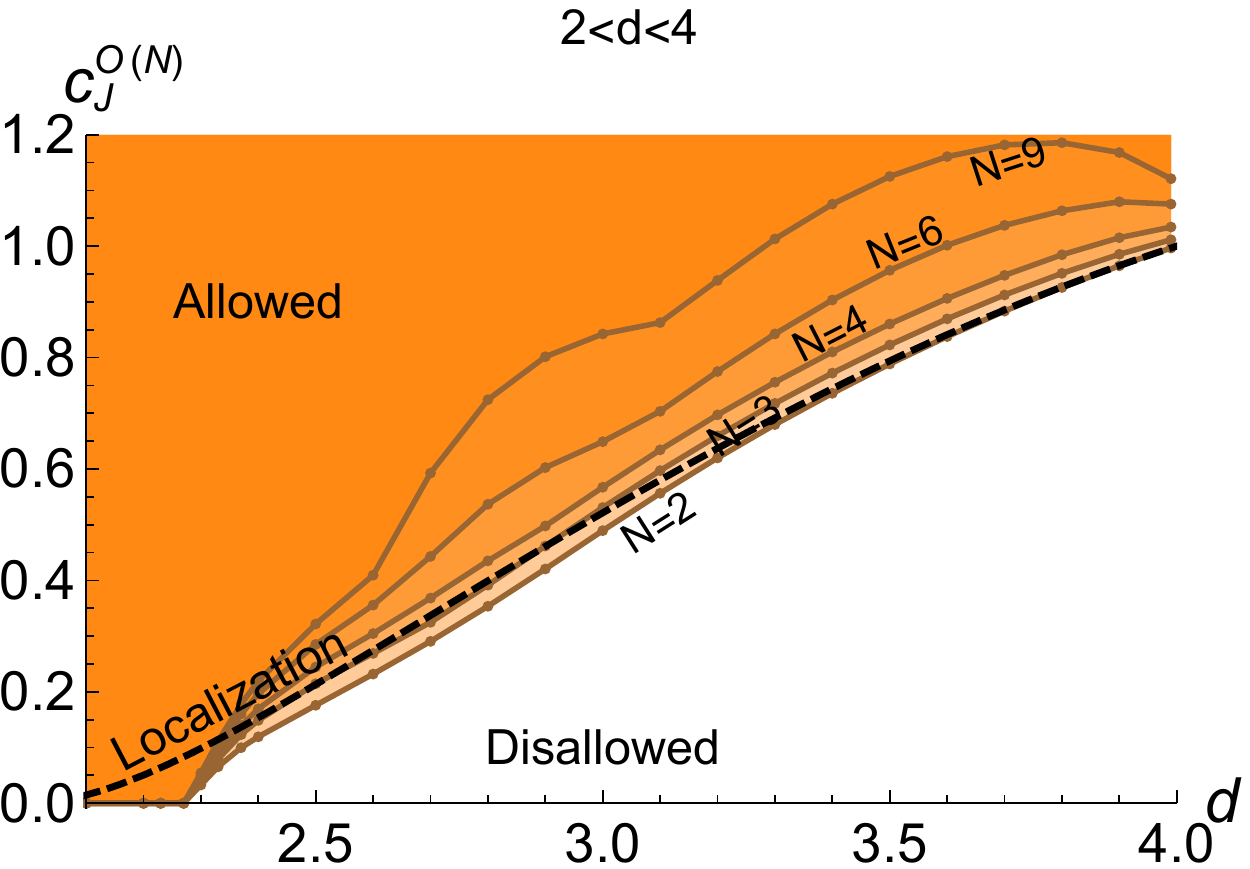}
  \includegraphics[width = 0.5\textwidth]{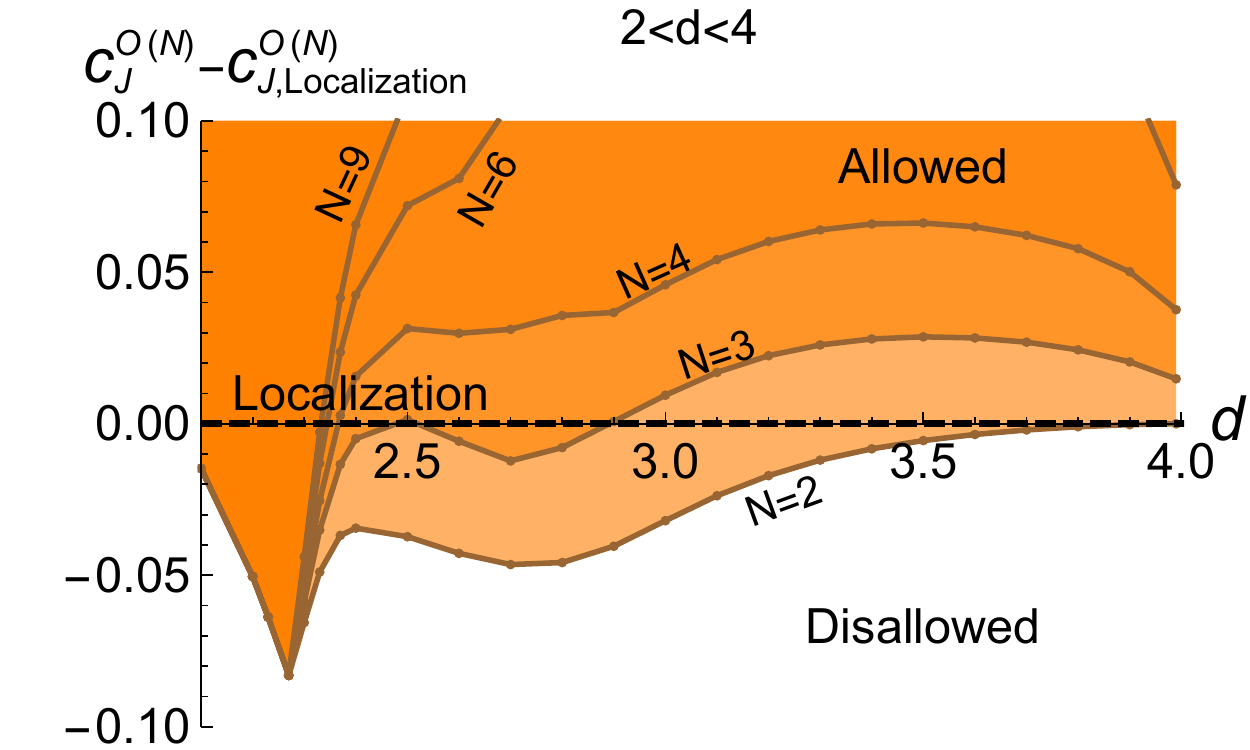}
 \caption{The left plot shows lower bounds on central charge $c_{J}^{O(N)}$  for $\mathcal{N}=2$ SCFTs with $O(N)$ symmetry in dimensions $2 < d<4$ for $N = 2,\, 3,\, 4,\, 6,\, 9$. The black dotted curve denotes the $N$ independent analytical value of $c_{J}^{O(N)}$ computed  from localization for SCFTs with super potential \eqref{SuperPot} in $2<d<4$. The right plot shows the difference between the $c_J^{O(N)}$ bound determined by the bootstrap and the results from localization, focusing on $N=2,\,3,\,4$. }
 \label{fig:allD}
 \end{center}
 \end{figure}

  Figure \ref{fig:cJvsDeltaZ} shows lower bounds on $c_J^{O(N)}$ as we vary $\Delta_{Z_i}$ for the specific case $N=3$ and $d=3$, together with the analytical value $(\Delta_{Z_i},c_J^{O(N)})=(2/3,0.521)$ determined by R-symmetry and localization. This value is marked with a black diamond in Figure~\ref{fig:cJvsDeltaZ} and is disallowed by the bootstrap bounds. As a consistency check, the plot begins at the free theory value $(\Delta_{Z_i},c_J^{O(N)})=(1/2,1)$, which lies on the boundary of the allowed region.  Also close to the boundary of the allowed region we find the fixed points with $O(N) \times U(1)$ flavor symmetry for which the values of $\Delta_{Z_i} = r_X$ and $c_J^{O(N)}$ were given in Table~\ref{RCharges}.  These latter SCFTs are marked with blue triangles in Figure~\ref{fig:cJvsDeltaZ}.  The dashed blue line that passes through these blue triangles represents the curve in the $(\Delta_Z, c_J^{O(N)})$ plane along which all these models are located, obtained by eliminating $N$ between the expressions for $\Delta_Z$ and $c_J^{O(N)}$.  As we can see, these $O(N) \times U(1)$-invariant SCFTs come close to saturating the bootstrap bounds, especially at large $N$ where the quantities $(\Delta_Z, c_J^{O(N)})$ approach the free field values $(1/2, 1)$.

In Figure \ref{fig:allD}, we depart from $d=3$ and show lower bounds on $c_J^{O(N)}$ at the value $\Delta_{Z_i}=\frac{d-1}{3}$ for $N=2,3,4$ in dimensions $2<d<4$, along with the localization value of $c_J^{O(N)}$ plotted as a function of dimension. In dimensions $d\geq3$ the localization value is disallowed by the bootstrap bounds for $N\geq3$, but as the dimension decreases toward $d=2$, theories with higher values of $N$ appear to be allowed by the bootstrap. However, the lower bounds computed using the bootstrap could conceivably be improved by inputing more theory specific assumptions into the bootstrap algorithm, such as the scaling dimensions of certain low-lying operators. The bounds on $N$ from the Generalized $F$-Theorem, discussed in Section~  \ref{Farguments}, are so far more stringent in $d<3$ than the bounds from the conformal bootstrap.

\section{Discussion}

In this paper, we argued that the ${\cal N} = 2$ generalization of the critical $O(N)$ vector model with superpotential \eqref{SuperPot} and $N>2$ exhibits a flavor symmetry enhancement at the infrared superconformal fixed point, where the coupling $g_2$ flows to zero.  While we present several arguments relying on a general RG flow analysis, the $4-\epsilon$ expansion, and the $F$-theorem in Section~\ref{INTUITION}, our most constraining argument is presented in Section~\ref{BOOTSTRAP} and relies on a combination between supersymmetric localization techniques and the conformal bootstrap.  It would be interesting to see if there are other situations in which the conformal bootstrap can be used in a similar way to provide an argument for symmetry enhancement.

In obtaining our bounds in Figure~\ref{fig:cJvsDeltaZ} for the $O(N)$ current central charge, we noticed that the superconformal fixed points with enhanced $O(N) \times U(1)$ global symmetry given in \eqref{SuperPotEnhanced} come close to saturating these bounds, being located at certain kinks in the boundary between the allowed and disallowed regions.  We hope to report on a more careful study of these SCFTs in a later publication \cite{LongerPaper}.

\section*{Acknowledgements}

We thank Matt Buican, Henriette Elvang, Simeon Hellerman, Daniel Jafferis, Ami Katz, David Simmons-Duffin, Edward Witten, and Kazuya Yonekura for useful discussions.  This work was supported in part by the US NSF under Grant No.~PHY-1418069 (SMC and SSP), PHY-1318681 (SG), and PHY-1314198 (IRK and RY)\@.  SSP thanks the Aspen Center for Physics for hospitality and partial support through NSF Grant No.~PHY-1066293 during the completion of this work.

\appendix

\section{Supersymmetric Localization}
\label{LOCALIZATION}

For the purpose of computing $c_J^{O(N)}$ using supersymmetric localization, it is useful to consider a $U(1)$ subgroup of $O(N)$ (call it $\widetilde{U(1)}$) that corresponds to $SO(2)$ rotations of the first two components of the $O(N)$ fundamental vector.  The conserved current is $\widetilde{j}^\mu = j^\mu_{12}$.  From \eqref{ONcurrent}, it obeys
 \es{cJU1t}{
 \langle \widetilde{j}^\mu (x) \widetilde{j}^\nu (0) \rangle =c_J^{O(N)}\frac{\Gamma^2(d/2)}{4(d-1)(d-2)\pi^d} 
      \left(\eta^{\mu\nu}-2\frac{x^\mu x^\nu}{x^2}\right)\frac{1}{x^{2d-2}} \,.
 }
In general, for an abelian current it is convenient to define 
 \es{cJU1tUsual}{
 \langle \widetilde{j}^\mu (x) \widetilde{j}^\nu (0) \rangle =c_J^{\widetilde{U(1)}}\frac{\Gamma^2(d/2)}{8(d-1)(d-2)\pi^d} 
      \left(\eta^{\mu\nu}-2\frac{x^\mu x^\nu}{x^2}\right)\frac{1}{x^{2d-2}} \,,
 }
so that a free chiral multiplet with unit charge has $c_J^{\widetilde{U(1)}} = 1$.    From comparing \eqref{cJU1t} to \eqref{cJU1tUsual}, we see that 
 \es{GotcJU1t}{
  c_J^{\widetilde{U(1)}} = 2c_J^{O(N)} \,.
 }

 We can now compute $c_J^{\widetilde{U(1)}}$ from supersymmetric localization.  Let us define $Y_{\pm} = Z_1 \pm i Z_2$ and write the superpotential \eqref{SuperPot} as
  \es{superpotAgain}{
  W = \frac{g_1}{2} \left(X Y_+ Y_- + \sum_{i = 3}^N X Z_i Z_i\right)
   + \frac{g_2}{6}  X^3 \,.
 }
Under $\widetilde{U(1)}$, $Y_\pm$ have charges $\pm 1$ while all other fields are neutral, so we consider the trial R-charges
 \es{TrialRCharges}{
  r_{Y_+} &= r_{Z^*}  +  y \,, \\
  r_{Y_-} &= r_{Z^*}  -   y \,, \\
  r_{Z_i} &= r_{Z^*} \,,
 }
where $y$ is the trial parameter and $r_{Z^*}$=$\frac{2}{3}$ at a fixed point of \eqref{SuperPot} with only $O(N) \times \Z_3$ flavor symmetry. In order to study the fixed point of \eqref{SuperPotEnhanced}, one should determine $r_{Z^*}$ by maximizing the $S^d$ free energy with respect to $r_{Z^*}$.   Eq.~\eqref{TrialRCharges} is consistent with the marginality of the superpotential \eqref{superpotAgain} and the coefficients $y$ are precisely equal to the $\widetilde{U(1)}$ charge.

The $S^d$ free energy is
 \es{S3Free}{
  F =  \tilde{\mathcal{F}}(\Delta_X)  +\tilde{\mathcal{F}}(\Delta_{Y_+})+\tilde{\mathcal{F}}(\Delta_{Y_-})+ \sum_{i = 3}^N \tilde{\mathcal{F}}(\Delta_{Z_i}) \,,
 }
 where $\tilde{\mathcal{F}}$ is the quantity defined in \cite{Giombi:2014xxa} that interpolates between the familiar Free Energies defined in integer dimensions. Here, we used $\Delta = (d-1) r /2$ as the relation between the scaling dimension and R-charge for a chiral operator.   For a single chiral super-field, $\widetilde{\mathcal{F}}$ can be compactly defined by its derivative:
\es{fracF}{
 \frac{\partial \widetilde{\mathcal{F}}(\Delta)}{\partial \Delta}=\frac{\Gamma(d-1-\Delta)\Gamma(\Delta)\sin(\pi(\Delta-d/2))}{\Gamma(d-1)} \,.
}

Regardless of $r_{Z^*}$, $F$ is maximized when $y = 0$.  The coefficient $c_J^{\widetilde{U(1)}}$ can be computed according to the results of \cite{Closset:2012vg} as
 \es{U1cJ}{
   c_J^{\widetilde{U(1)}}= -\left(\frac{2^d \Gamma( \frac{d-1}{2})}{(d-1)^2 \pi^{3/2} \Gamma(\frac d2 - 1)}  \right)\frac{\partial^2 F}{\partial y^2}\bigg |_{y=0} \,,
 }
 where the constant of proportionality is fixed by requiring that $c_J^{\widetilde{U(1)}}$ equal one for the theory of a single free chiral multiplet.

Combining \eqref{GotcJU1t},\eqref{S3Free}, \eqref{fracF}, and \eqref{U1cJ}, and using $r_{Z^*} = 2/3$ for the proposed fixed point of \eqref{SuperPot} with $O(N) \times \Z_3$ flavor symmetry, we can calculate $c_J^{O(N)}$ for arbitrary $d$.  See Table~\ref{cJONTable} for numerical results in various $d$.

If we wish to compute $c_J^{O(N)}$ for the fixed point of \eqref{SuperPotEnhanced} with $O(N) \times U(1)$ flavor symmetry, we should first maximize $F$ with respect to $r_{Z^*}$.  In $d=3$, this maximization was performed in \cite{Nishioka:2013gza} and the result is also listed in Table~\ref{RCharges}.  In Table~\ref{RCharges} we also list the corresponding values of $c_J^{O(N)}$ computed from \eqref{GotcJU1t} and \eqref{U1cJ}.

\section{$\mathcal{N}=2$ $O(N)$ Conformal Bootstrap}
\label{CROSSING}

We now show how to constrain $\cN = 2$ theories with $O(N)$ symmetry using the conformal bootstrap technique.  We restrict our attention to the four-point function of two chiral and two anti-chiral scalar operators transforming in the fundamental representation of $O(N)$.  Let $Z_i$ be such a chiral operator, with $i = 1, \ldots, N$ an $O(N)$ fundamental index, and let $\bar Z_i$ be an anti-chiral operator.  We find it convenient to write $Z_{i} = Z_{1i} + i Z_{2i}$ and $\bar Z_{i} = Z_{1i} - i Z_{2i}$ and work with $Z_{Ii}$, ($I= 1, 2$ being a fundamental $SO(2)_R$ index) instead of $Z_i$ and $\bar Z_i$.    We will examine the four-point function:
\es{4pt}{
  \<  Z_{I i} (x_1)Z_{J j}(x_2) Z_{K k}(x_3) Z_{L l}(x_4) \> \,,
}
which includes all orderings of two $Z$'s and two $\bar Z$'s at once.

The operators appearing in the $Z_{I i} \times Z_{J j}$ OPE can be classified according to their transformation properties under $SO(2)_R \times O(N)$.  We have singlets or rank-two traceless symmetric tensors of $SO(2)_R$ (corresponding to operators that have zero or $\pm 2 r_Z$ R-charge, respectively) denoted by $S$ and $T$, as well as singlets, rank-two traceless symmetric tensors, or rank-two anti-symmetric tenors of $O(N)$ denoted by $s$, $t$, and $a$, respectively.  Due to Bose symmetry, the operators of the type $Ts$ and $Tt$ must have even spin, those of the type $Ta$ should have odd spin, and there are no spin restrictions on the other operators.

Forgetting about supersymmetry and performing the $s$-channel OPE in \eqref{4pt}, one can write the four-point function as a sum over conformal blocks $G_{\Delta, \ell}$ for identical scalar operators \cite{Dolan:2011dv}.  We have\footnote{We use the normalization of the conformal blocks in \cite{Kos:2013tga}.  Specifically, in the $r$ and $\eta$ coordinates introduced in \cite{Hogervorst:2013sma}, we have $G_{\Delta, \ell} = r^\Delta P_\ell(\eta) + \ldots$, as $r \to 0$ with $\eta$ kept fixed.}
\es{crossing}{
&x_{12}^{2\Delta_Z}  x_{34}^{2\Delta_Z} \<  Z_{I i} (x_1)Z_{J j}(x_2) Z_{K k}(x_3) Z_{L l}(x_4) \> \\
  &= F_{IJKL}^{(1)} \left[ \sum_{\substack{\cO \in Ss \\ \ell \ \text{even}}} f_{ijkl}^{(1)}  \lambda_\mathcal{O}^2
   G_{\Delta, \ell} (u, v)
    + \sum_{\substack{\cO \in Ss \\ \ell \ \text{odd}}}  f_{ijkl}^{(2)}   \lambda_\mathcal{O}^2
   G_{\Delta, \ell} (u, v)
    + \sum_{\substack{\cO \in Ts \\ \ell \ \text{even}}}  f_{ijkl}^{(3)}  \lambda_\mathcal{O}^2
   G_{\Delta, \ell} (u, v)\right] \\
  &+ F_{IJKL}^{(2)} \left[ \sum_{\substack{\cO \in Sa \\ \ell \ \text{odd}}} f_{ijkl}^{(1)}  \lambda_\mathcal{O}^2
   G_{\Delta, \ell} (u, v)
    + \sum_{\substack{\cO \in Sa \\ \ell \ \text{even}}}  f_{ijkl}^{(2)}   \lambda_\mathcal{O}^2
   G_{\Delta, \ell} (u, v)
    + \sum_{\substack{\cO \in Ta \\ \ell \ \text{odd}}}  f_{ijkl}^{(3)}  \lambda_\mathcal{O}^2
   G_{\Delta, \ell} (u, v)\right] \\
  &+ F_{IJKL}^{(3)} \left[ \sum_{\substack{\cO \in St \\ \ell \ \text{even}}} f_{ijkl}^{(1)}  \lambda_\mathcal{O}^2
   G_{\Delta, \ell} (u, v)
    + \sum_{\substack{\cO \in St \\ \ell \ \text{odd}}}  f_{ijkl}^{(2)}   \lambda_\mathcal{O}^2
   G_{\Delta, \ell} (u, v)
    + \sum_{\substack{\cO \in Tt \\ \ell \ \text{even}}}  f_{ijkl}^{(3)}  \lambda_\mathcal{O}^2
   G_{\Delta, \ell} (u, v)\right]   \,,
}
where
 \es{fDefs}{
  f_{ijkl}^{(1)} = \delta_{ij} \delta_{kl} \,, \qquad
   f_{ijkl}^{(2)} =  \delta_{il} \delta_{jk} -  \delta_{ik} \delta_{jl} \,, \qquad
    f_{ijkl}^{(3)} = \delta_{il} \delta_{jk} +  \delta_{ik} \delta_{jl}  - \delta_{ij} \delta_{kl} \,,
 }
and
   \es{FDefs}{
  F_{IJKL}^{(1)}  &=   \delta_{IJ} \delta_{KL} \,, \qquad
   F_{IJKL}^{(2)} = \delta_{IL} \delta_{JK} - \delta_{IK} \delta_{JL} \,, \\
    F_{IJKL}^{(3)} &= \delta_{IL} \delta_{JK} - \delta_{IK} \delta_{JL}  - \frac{2}{N} \delta_{IJ} \delta_{KL} \,.
 }
In the sum \eqref{crossing}, we sum only over conformal primaries ${\cal O}$.

Supersymmetry relates some of the OPE coefficients in \eqref{crossing} to one another.  As explained in \cite{Bobev:2015jxa}, in the $T$ channel, there is only one operator per superconformal multiplet contributing to \eqref{crossing}.  In the $S$ channel, there are generically four operators contributing that have related OPE coefficients.  Their contributions can be grouped into a superconformal block.  As in \cite{Bobev:2015jxa}, let us define
 \es{SuperConf}{
\cG_{\Delta,\ell} &= G_{\Delta ,\ell}  +\frac{2(\ell+d-2)(\Delta +\ell) }{(2\ell+d-2)(\Delta +\ell+1)} G_{\Delta +1,\ell+1} \\
&+ \frac{2\ell(\ell+d-3)(2\ell+d-4) (\Delta - \ell+2-d)}{ ( \ell+d-3)(2\ell+d-4)(2\ell+d-2) (\Delta-\ell-d+3 )}G_{\Delta +1,\ell-1} \\
&+\frac{\Delta(\Delta+3-d) (\Delta - \ell +2-d) (\Delta +\ell) }{\left( \Delta+2-\frac d 2 \right) \left(\Delta+1-\frac d 2 \right)(\Delta-\ell+3-d ) (\Delta +\ell+1)} G_{\Delta +2,\ell} \,.
 }
We also define $\tilde \cG_{\Delta, \ell}$ to be the same expression as \eqref{SuperConf} with the middle two terms multiplied by $(-1)$.  Then, combining the contributions coming from the same superconformal multiplet amounts to the replacements
 \es{Replacements}{
  f_{ijkl}^{(1)} G_{\Delta, \ell} &\to f_{ijkl}^{(1)} \frac{ \cG_{\Delta, \ell}  + \tilde \cG_{\Delta, \ell} }{2} + f_{ijkl}^{(2)} \frac{ \cG_{\Delta, \ell}  - \tilde \cG_{\Delta, \ell} }{2} \,,\\
   f_{ijkl}^{(2)} G_{\Delta, \ell} &\to f_{ijkl}^{(2)} \frac{ \cG_{\Delta, \ell}  + \tilde \cG_{\Delta, \ell} }{2} + f_{ijkl}^{(1)} \frac{ \cG_{\Delta, \ell}  - \tilde \cG_{\Delta, \ell} }{2}
 }
in \eqref{crossing}.  With these replacements, we should sum only over superconformal primaries in the first two terms of each line of \eqref{crossing}.


The four-point function \eqref{crossing} (with or without the replacements \eqref{Replacements}) should be invariant under crossing symmetry, whereby one exchanges two of the operators.  Some of these exchanges are trivial---for instance, the $\{1, I, i\} \leftrightarrow \{2, J, j\}$ exchange yields nothing but the selection rules on the spins of the operators that appear in each channel in \eqref{crossing}.  The $\{1, I, i\} \leftrightarrow \{3, K, k\}$ exchange yields non-trivial conditions, given by $9$ equations, which can be grouped into a vector ``sum rule.''  Upon using \eqref{Replacements}, we obtain
 \es{eq:crossingWithON}{
   0 &= \sum_{Ss, \text{ all }\ell}  \lambda_\mathcal{O}^2 V^{Ss}_{ \Delta, \ell}
      + \sum_{{St, \text{ all }\ell }}  \lambda_\mathcal{O}^2 V^{St}_{\Delta, \ell}
      + \sum_{{Sa, \text{ all }\ell }}  \lambda_\mathcal{O}^2 V^{Sa}_{ \Delta, \ell} \\
     &{}+  \sum_{\substack{Ts, \text{ }\ell \text{ even }}}  \lambda_\mathcal{O}^2 V^{Ts}_{\Delta, \ell}
       + \sum_{\substack{Tt, \text{ }\ell \text{ even }}}     \lambda_\mathcal{O}^2 V^{Tt}_{ \Delta, \ell}
       +  \sum_{\substack{Ta, \text{ }\ell \text{ odd }}}   \lambda_\mathcal{O}^2 V^{Ta}_{\Delta, \ell} \,,
 }
where the  $V_{\Delta, \ell}$ are given by
 \es{VDefs}{
V_{ \Delta, \ell}^{Rs} = \left(
\begin{matrix}
0\\
U^{-,R}_{\Delta,\ell}\\
U^{+, R}_{\Delta,\ell}
\end{matrix} \right) \,, \qquad  V_{ \Delta, \ell}^{Rt} = \left(
\begin{matrix}
U^{-, R}_{\Delta,\ell}\\
\left(1-\dfrac{2}{N}\right)U^{-, R}_{\Delta,\ell}\\
-\left(1+\dfrac{2}{N}\right)U^{+, R}_{\Delta,\ell}
\end{matrix} \right) \,, \qquad V_{\Delta, \ell}^{Ra} = \left(
\begin{matrix}
-U^{-, R}_{\Delta,\ell}\\
U^{-, R}_{\Delta,\ell}\\
-U^{+, R}_{\Delta,\ell}
\end{matrix} \right),
 }
for which $R \in \{S,  T\}$, and the $U^{\pm, R}_{\Delta, \ell}$ are given by
 \es{UDefs}{
U^{\pm, S}_{\Delta,\ell} = \left(\begin{matrix}
\mathcal{F}_{\Delta, \ell}^\mp\\
\mathcal{\tilde F}_{\Delta, \ell}^\mp\\
\mathcal{\tilde F}_{\Delta, \ell}^\pm
\end{matrix}\right) \,, \qquad  U^{\pm, T}_{\Delta,\ell} = \left(\begin{matrix}
0\\
{F}_{\Delta, \ell}^\mp\\
-{F}_{\Delta, \ell}^\pm
\end{matrix}\right) \,.
 }
Here, we defined
 \es{eq:defOfF}{
   {F}_{\Delta,\ell}^{\pm}  &= v^{\Delta_Z}  G_{\Delta, \ell}(u, v) \pm u^{\Delta_Z}  G_{\Delta, \ell}(v, u) \,, \\
   \mathcal{F}_{\Delta,\ell}^{\pm}  &= v^{\Delta_Z} \mathcal G_{\Delta, \ell}(u, v) \pm u^{\Delta_Z} \mathcal G_{\Delta, \ell}(v, u)\,, \\
  \tilde{\mathcal{F}}_{\Delta,\ell}^{\pm}  &= v^{\Delta_Z} \tilde{\mathcal G}_{\Delta, \ell}(u, v) \pm u^{\Delta_Z} \tilde{\mathcal G}_{\Delta, \ell}(v, u)\,.
 }

The operator spectrum is further constrained due to the $\mathcal{N}=2$ supersymmetry \cite{Bobev:2015vsa,Bobev:2015jxa}. Specifically, by generalizing the reasoning in \cite{Bobev:2015jxa} for $\cN=2$ SCFTs to include $O(N)$ symmetry, we find the constraints listed in Table~\ref{UnitarityTable}.

\begin{table}[htpb]
\centering
\begin{tabular}{c|c|c|c|}
\cline{2-4}
& \hspace{0.5in} $s$ \hspace{0.5in} & \hspace{0.5in} $t$ \hspace{0.5in}  & $a$  \\
\hline
\multicolumn{1}{|c|}{$S$} & \multicolumn{3}{c|}{$\Delta \geq \ell + d - 2$,\  \ for all allowed values of $\ell$} \\
\hline
\multicolumn{1}{|c|}{\multirow{4}{*}{$T$}} & \multicolumn{3}{c|}{$\Delta \geq \abs{2 \Delta_Z - (d-1) } + \ell + (d - 1)$, \ for all allowed values of $\ell$} \\
\multicolumn{1}{|c|}{} & \multicolumn{3}{c|}{$\Delta =2 \Delta_Z + \ell$, \ for all allowed values of $\ell$} \\
\cline{2-4}
\multicolumn{1}{|c|}{}  & \multicolumn{2}{c|}{$\Delta = d - 2 \Delta_Z$, \ for $\ell = 0$, $\Delta_Z \leq d/4$} & \\
\multicolumn{1}{|c|}{}  & \multicolumn{2}{c|}{$\Delta = 2 \Delta_Z$, \ for $\ell = 0$} & \\
\hline
\end{tabular}
\caption{Constraints on the operator spectrum of an ${\cal N} = 2$ SCFT with $O(N)$ flavor symmetry coming from supersymmetry.}\label{UnitarityTable}
\end{table}

We are interested in bounding $c_J^{O(N)}$. This quantity is related to the OPE coefficient $\lambda_{Sa,d-1,1}$ with which the $O(N)$ current, $j^\mu_{ij}$, appears in the $Z_i \times \bar Z_j$ OPE\@.  Since $j^\mu_{ij}$ is a superconformal descendant, its OPE coefficient $\lambda_{Sa,d-1,1}$ is related to the OPE coefficient $\lambda_{Sa,d-2,0}$ of the $O(N)$-antisymmetric spin-$0$  superconformal primary $\cO_{Sa,d-2,0}$ in the same supermultiplet as $j^\mu_{ij}$.  According to \eqref{SuperConf}, we can relate $\lambda^2_{Sa,d-1,1}$ to $\lambda^2_{Sa,d-2,0}$ as
\es{superRel}{
\lambda^2_{Sa,d-1,1}=\lambda^2_{Sa,d-2,0}\frac{2(d-2)}{d-1} \,.
}
By expanding the four point function \eqref{4pt} of the free theory in terms of superconformal blocks, we can derive the relationship
\es{freeRel}{
c_J^{O(N)}=\lambda^2_{Sa,d-1,1}\frac{4^{d-2}(d-2)}{d-1} \,.
}
Combining \eqref{superRel} and \eqref{freeRel}, we find
  \es{ONcJLambda2}{
 c_J^{O(N)}=\frac{2^{2d-5}}{\lambda^2_{Sa,d-2, 0}} \,.
 }
To bound $\lambda^2_{Sa,d-2, 0}$, we start by rewriting the crossing equation \eqref{crossing} as \cite{El-Showk:2014dwa},
\es{crossing2}{
\lambda^2_{Sa,d-2, 0} V_{d-2,0}^{Sa} = -V^{Ss}_{\text{unit}} - \sum_{\cO \neq \cO_{Sa,d-2, 0}} \lambda^2_\cO V_\cO \,.
}
Now apply a linear functional $\vec \alpha$ to \eqref{crossing2} and look at the space of functionals that satisfy the constraints
\es{alphas}{
\vec\alpha\left(\vec V_{d-2,0}^{Sa}\right) =& 1 \,, \\
\vec\alpha \left( \vec V_{\cO}(\Delta)\right) \geq& 0 \text{ for all $\cO\neq \cO_{Sa,0}$ and constraining $\Delta$ as in Table~\ref{UnitarityTable}. }
}
Eqs.~\eqref{crossing2} and \eqref{alphas} then imply that
\es{finalCross}{
 \lambda^{2}_{Sa,d-2, 0} \leq  \vec \alpha(- \vec V^{Ss}_{\text{unit}}) \,.
}
By finding the minimal such $\vec\alpha$ we find an upper bound on $\lambda^{2}_{Sa, d-2, 0}$, which using \eqref{ONcJLambda2} gives a lower bound on $c_J^{O(N)}$ .

The numerical results presented in the main text were generated as follows.  We used a \texttt{Mathematica} script to generate the conformal blocks $G_{\Delta, \ell}(u, v)$ in arbitrary dimension, using the recursion formula for scalar conformal blocks \cite{Kos:2014bka}. We implemented the semi-definite programming required by the numerical bootstrap using \texttt{SDPB} \cite{Simmons-Duffin:2015qma}, for which we used the parameters specified in the first column of Table 1 in the \texttt{SDPB} manual \cite{Simmons-Duffin:2015qma}. The convergence of our results was tested by varying the maximum number of derivatives, $\Lambda$, of the functionals $\vec\alpha$---specifically, we notice that the bound for $\Lambda=19$ differ from those with $\Lambda=21$ by $10^{-3}$ in the value of $c_J^{O(N)}$. Furthermore, as we increase the search space in $\vec{\alpha}$, the space of allowed theories can only become smaller, which implies that once a theory is excluded by the numerical bootstrap at a given value of $\Lambda$, it is rigorously excluded from the space of all mathematically consistent unitary CFTs.

\bibliographystyle{ssg}
\bibliography{cubic}

\begingroup\raggedright\begin{thebibliography}{10}

\bibitem{Jafferis:2010un}
D.~L. Jafferis, ``{The Exact Superconformal R-Symmetry Extremizes $Z$},'' {\em
  JHEP} {\bf 1205} (2012) 159, \href{http://xxx.lanl.gov/abs/1012.3210}{{\tt
  1012.3210}}.

\bibitem{Jafferis:2011zi}
D.~L. Jafferis, I.~R. Klebanov, S.~S. Pufu, and B.~R. Safdi, ``{Towards the
  $F$-Theorem: ${\cal N}=2$ Field Theories on the Three-Sphere},'' {\em JHEP}
  {\bf 1106} (2011) 102, \href{http://xxx.lanl.gov/abs/1103.1181}{{\tt
  1103.1181}}.

\bibitem{Closset:2012vg}
C.~Closset, T.~T. Dumitrescu, G.~Festuccia, Z.~Komargodski, and N.~Seiberg,
  ``{Contact Terms, Unitarity, and $F$-Maximization in Three-Dimensional
  Superconformal Theories},'' {\em JHEP} {\bf 1210} (2012) 053,
  \href{http://xxx.lanl.gov/abs/1205.4142}{{\tt 1205.4142}}.

\bibitem{Minwalla:1997ka}
S.~Minwalla, ``{Restrictions imposed by superconformal invariance on quantum
  field theories},'' {\em Adv.Theor.Math.Phys.} {\bf 2} (1998) 781--846,
  \href{http://xxx.lanl.gov/abs/hep-th/9712074}{{\tt hep-th/9712074}}.

\bibitem{Hama:2010av}
N.~Hama, K.~Hosomichi, and S.~Lee, ``{Notes on SUSY Gauge Theories on
  Three-Sphere},'' {\em JHEP} {\bf 1103} (2011) 127,
  \href{http://xxx.lanl.gov/abs/1012.3512}{{\tt 1012.3512}}.

\bibitem{Niarchos:2011sn}
V.~Niarchos, ``{Comments on F-maximization and R-symmetry in 3D SCFTs},'' {\em
  J. Phys.} {\bf A44} (2011) 305404,
  \href{http://xxx.lanl.gov/abs/1103.5909}{{\tt 1103.5909}}.

\bibitem{Morita:2011cs}
T.~Morita and V.~Niarchos, ``{$F$-theorem, duality and SUSY breaking in
  one-adjoint Chern-Simons-Matter theories},'' {\em Nucl. Phys.} {\bf B858}
  (2012) 84--116, \href{http://xxx.lanl.gov/abs/1108.4963}{{\tt 1108.4963}}.

\bibitem{Agarwal:2012wd}
P.~Agarwal, A.~Amariti, and M.~Siani, ``{Refined Checks and Exact Dualities in
  Three Dimensions},'' {\em JHEP} {\bf 10} (2012) 178,
  \href{http://xxx.lanl.gov/abs/1205.6798}{{\tt 1205.6798}}.

\bibitem{Kutasov:2003iy}
D.~Kutasov, A.~Parnachev, and D.~A. Sahakyan, ``{Central charges and $U(1)_R$
  symmetries in ${\cal N}=1$ super-Yang-Mills},'' {\em JHEP} {\bf 0311} (2003)
  013, \href{http://xxx.lanl.gov/abs/hep-th/0308071}{{\tt hep-th/0308071}}.

\bibitem{Safdi:2012re}
B.~R. Safdi, I.~R. Klebanov, and J.~Lee, ``{A Crack in the Conformal Window},''
  {\em JHEP} {\bf 1304} (2013) 165,
  \href{http://xxx.lanl.gov/abs/1212.4502}{{\tt 1212.4502}}.

\bibitem{Myers:2010xs}
R.~C. Myers and A.~Sinha, ``{Seeing a $c$-theorem with holography},'' {\em
  Phys.Rev.} {\bf D82} (2010) 046006,
  \href{http://xxx.lanl.gov/abs/1006.1263}{{\tt 1006.1263}}.

\bibitem{Klebanov:2011gs}
I.~R. Klebanov, S.~S. Pufu, and B.~R. Safdi, ``{$F$-Theorem without
  Supersymmetry},'' {\em JHEP} {\bf 1110} (2011) 038,
  \href{http://xxx.lanl.gov/abs/1105.4598}{{\tt 1105.4598}}.

\bibitem{Casini:2012ei}
H.~Casini and M.~Huerta, ``{On the RG running of the entanglement entropy of a
  circle},'' {\em Phys.Rev.} {\bf D85} (2012) 125016,
  \href{http://xxx.lanl.gov/abs/1202.5650}{{\tt 1202.5650}}.

\bibitem{Rattazzi:2008pe}
R.~Rattazzi, V.~S. Rychkov, E.~Tonni, and A.~Vichi, ``{Bounding scalar operator
  dimensions in 4D CFT},'' {\em JHEP} {\bf 0812} (2008) 031,
  \href{http://xxx.lanl.gov/abs/0807.0004}{{\tt 0807.0004}}.

\bibitem{Poland:2010wg}
D.~Poland and D.~Simmons-Duffin, ``{Bounds on 4D Conformal and Superconformal
  Field Theories},'' {\em JHEP} {\bf 1105} (2011) 017,
  \href{http://xxx.lanl.gov/abs/1009.2087}{{\tt 1009.2087}}.

\bibitem{Green:2010da}
D.~Green, Z.~Komargodski, N.~Seiberg, Y.~Tachikawa, and B.~Wecht, ``{Exactly
  Marginal Deformations and Global Symmetries},'' {\em JHEP} {\bf 1006} (2010)
  106, \href{http://xxx.lanl.gov/abs/1005.3546}{{\tt 1005.3546}}.

\bibitem{Strassler:1998iz}
M.~J. Strassler, ``{On renormalization group flows and exactly marginal
  operators in three-dimensions},''
  \href{http://xxx.lanl.gov/abs/hep-th/9810223}{{\tt hep-th/9810223}}.

\bibitem{Bobev:2015jxa}
N.~Bobev, S.~El-Showk, D.~Mazac, and M.~F. Paulos, ``{Bootstrapping SCFTs with
  Four Supercharges},'' \href{http://xxx.lanl.gov/abs/1503.02081}{{\tt
  1503.02081}}.

\bibitem{Martelli:2009ga}
D.~Martelli and J.~Sparks, ``{$AdS_4 /CFT_3$ duals from M2-branes at
  hypersurface singularities and their deformations},'' {\em JHEP} {\bf 12}
  (2009) 017, \href{http://xxx.lanl.gov/abs/0909.2036}{{\tt 0909.2036}}.

\bibitem{Strassler:2003qg}
M.~J. Strassler, ``{An Unorthodox introduction to supersymmetric gauge
  theory},'' in {\em {Strings, Branes and Extra Dimensions: TASI 2001:
  Proceedings}}, pp.~561--638, 2003.
\newblock \href{http://xxx.lanl.gov/abs/hep-th/0309149}{{\tt hep-th/0309149}}.

\bibitem{Nishioka:2013gza}
T.~Nishioka and K.~Yonekura, ``{On RG Flow of $\tau_{rr}$ for Supersymmetric
  Field Theories in Three-Dimensions},'' {\em JHEP} {\bf 1305} (2013) 165,
  \href{http://xxx.lanl.gov/abs/1303.1522}{{\tt 1303.1522}}.

\bibitem{Ferreira:1996az}
P.~Ferreira, I.~Jack, and D.~Jones, ``{The Quasiinfrared fixed point at higher
  loops},'' {\em Phys.Lett.} {\bf B392} (1997) 376--382,
  \href{http://xxx.lanl.gov/abs/hep-ph/9610296}{{\tt hep-ph/9610296}}.

\bibitem{Giombi:2014xxa}
S.~Giombi and I.~R. Klebanov, ``{Interpolating between $a$ and $F$},'' {\em
  JHEP} {\bf 1503} (2015) 117, \href{http://xxx.lanl.gov/abs/1409.1937}{{\tt
  1409.1937}}.

\bibitem{Fei:2015oha}
L.~Fei, S.~Giombi, I.~R. Klebanov, and G.~Tarnopolsky, ``{Generalized
  $F$-Theorem and the $\epsilon$ Expansion},''
  \href{http://xxx.lanl.gov/abs/1507.01960}{{\tt 1507.01960}}.

\bibitem{Vafa:1988uu}
C.~Vafa and N.~P. Warner, ``{Catastrophes and the Classification of Conformal
  Theories},'' {\em Phys.Lett.} {\bf B218} (1989) 51.

\bibitem{Martinec:1988zu}
E.~J. Martinec, ``{Algebraic Geometry and Effective Lagrangians},'' {\em
  Phys.Lett.} {\bf B217} (1989) 431.

\bibitem{Rattazzi:2010yc}
R.~Rattazzi, S.~Rychkov, and A.~Vichi, ``{Bounds in 4D Conformal Field Theories
  with Global Symmetry},'' {\em J.Phys.} {\bf A44} (2011) 035402,
  \href{http://xxx.lanl.gov/abs/1009.5985}{{\tt 1009.5985}}.

\bibitem{ElShowk:2012ht}
S.~El-Showk, M.~F. Paulos, D.~Poland, S.~Rychkov, D.~Simmons-Duffin, {\em
  et.~al.}, ``{Solving the 3D Ising Model with the Conformal Bootstrap},'' {\em
  Phys.Rev.} {\bf D86} (2012) 025022,
  \href{http://xxx.lanl.gov/abs/1203.6064}{{\tt 1203.6064}}.

\bibitem{El-Showk:2014dwa}
S.~El-Showk, M.~F. Paulos, D.~Poland, S.~Rychkov, D.~Simmons-Duffin, {\em
  et.~al.}, ``{Solving the 3d Ising Model with the Conformal Bootstrap II.
  $c$-Minimization and Precise Critical Exponents},''
  \href{http://xxx.lanl.gov/abs/1403.4545}{{\tt 1403.4545}}.

\bibitem{Vichi:2011ux}
A.~Vichi, ``{Improved bounds for CFT's with global symmetries},'' {\em JHEP}
  {\bf 1201} (2012) 162, \href{http://xxx.lanl.gov/abs/1106.4037}{{\tt
  1106.4037}}.

\bibitem{Kos:2013tga}
F.~Kos, D.~Poland, and D.~Simmons-Duffin, ``{Bootstrapping the $O(N)$ Vector
  Models},'' \href{http://xxx.lanl.gov/abs/1307.6856}{{\tt 1307.6856}}.

\bibitem{Chester:2014gqa}
S.~M. Chester, S.~S. Pufu, and R.~Yacoby, ``{Bootstrapping $O(N)$ vector models
  in 4 $< d <$ 6},'' {\em Phys.Rev.} {\bf D91} (2015), no.~8 086014,
  \href{http://xxx.lanl.gov/abs/1412.7746}{{\tt 1412.7746}}.

\bibitem{Nakayama:2014lva}
Y.~Nakayama and T.~Ohtsuki, ``{Approaching conformal window of $O(n)\times
  O(m)$ symmetric Landau-Ginzburg models from conformal bootstrap},''
  \href{http://xxx.lanl.gov/abs/1404.0489}{{\tt 1404.0489}}.

\bibitem{Nakayama:2014sba}
Y.~Nakayama and T.~Ohtsuki, ``{Bootstrapping phase transitions in QCD and
  frustrated spin systems},'' \href{http://xxx.lanl.gov/abs/1407.6195}{{\tt
  1407.6195}}.

\bibitem{Poland:2011ey}
D.~Poland, D.~Simmons-Duffin, and A.~Vichi, ``{Carving Out the Space of 4D
  CFTs},'' {\em JHEP} {\bf 1205} (2012) 110,
  \href{http://xxx.lanl.gov/abs/1109.5176}{{\tt 1109.5176}}.

\bibitem{Beem:2013qxa}
C.~Beem, L.~Rastelli, and B.~C. van Rees, ``{The $\mathcal N=4$ Superconformal
  Bootstrap},'' {\em Phys.Rev.Lett.} {\bf 111} (2013) 071601,
  \href{http://xxx.lanl.gov/abs/1304.1803}{{\tt 1304.1803}}.

\bibitem{Beem:2013hha}
C.~Beem, L.~Rastelli, A.~Sen, and B.~C. van Rees, ``{Resummation and S-duality
  in ${\cal N}=4$ SYM},'' \href{http://xxx.lanl.gov/abs/1306.3228}{{\tt
  1306.3228}}.

\bibitem{Alday:2013opa}
L.~F. Alday and A.~Bissi, ``{The superconformal bootstrap for structure
  constants},'' \href{http://xxx.lanl.gov/abs/1310.3757}{{\tt 1310.3757}}.

\bibitem{Alday:2013bha}
L.~F. Alday and A.~Bissi, ``{Modular interpolating functions for ${\cal N}=4$
  SYM},'' \href{http://xxx.lanl.gov/abs/1311.3215}{{\tt 1311.3215}}.

\bibitem{Chester:2014fya}
S.~M. Chester, J.~Lee, S.~S. Pufu, and R.~Yacoby, ``{The $ \mathcal{N}=8 $
  superconformal bootstrap in three dimensions},'' {\em JHEP} {\bf 1409} (2014)
  143, \href{http://xxx.lanl.gov/abs/1406.4814}{{\tt 1406.4814}}.

\bibitem{Chester:2014mea}
S.~M. Chester, J.~Lee, S.~S. Pufu, and R.~Yacoby, ``{Exact Correlators of BPS
  Operators from the 3d Superconformal Bootstrap},'' {\em JHEP} {\bf 1503}
  (2015) 130, \href{http://xxx.lanl.gov/abs/1412.0334}{{\tt 1412.0334}}.

\bibitem{LongerPaper}
S.~Chester, S.~Giombi, L.~Iliesiu, I.~Klebanov, S.~Pufu, and R.~Yacoby,
  forthcoming.

\bibitem{Dolan:2011dv}
F.~Dolan and H.~Osborn, ``{Conformal Partial Waves: Further Mathematical
  Results},'' \href{http://xxx.lanl.gov/abs/1108.6194}{{\tt 1108.6194}}.

\bibitem{Hogervorst:2013sma}
M.~Hogervorst and S.~Rychkov, ``{Radial Coordinates for Conformal Blocks},''
  {\em Phys.Rev.} {\bf D87} (2013), no.~10 106004,
  \href{http://xxx.lanl.gov/abs/1303.1111}{{\tt 1303.1111}}.

\bibitem{Bobev:2015vsa}
N.~Bobev, S.~El-Showk, D.~Mazac, and M.~F. Paulos, ``{Bootstrapping the
  Three-Dimensional Supersymmetric Ising Model},''
  \href{http://xxx.lanl.gov/abs/1502.04124}{{\tt 1502.04124}}.

\bibitem{Kos:2014bka}
F.~Kos, D.~Poland, and D.~Simmons-Duffin, ``{Bootstrapping Mixed Correlators in
  the 3D Ising Model},'' \href{http://xxx.lanl.gov/abs/1406.4858}{{\tt
  1406.4858}}.

\bibitem{Simmons-Duffin:2015qma}
D.~Simmons-Duffin, ``{A Semidefinite Program Solver for the Conformal
  Bootstrap},'' \href{http://xxx.lanl.gov/abs/1502.02033}{{\tt 1502.02033}}.

\end{thebibliography}\endgroup

\end{document}